\RequirePackage{fix-cm}
\documentclass[twocolumn]{svjour3}          
\usepackage{graphicx}
\usepackage{placeins}
\usepackage{microtype}
\begin{document}

\title{Technologies for trapped-ion quantum information systems}
\subtitle{Progress towards scalability with hybrid systems}

\author{Amira M. Eltony \and Dorian Gangloff \and Molu Shi \and Alexei Bylinskii \and Vladan Vuleti\'{c} \and Isaac L. Chuang}

\institute{Center for Ultracold Atoms, Research Laboratory of Electronics, and Department of Physics, Massachusetts Institute of Technology, Cambridge, Massachusetts 02139, USA \\
\email{aeltony@mit.edu}
}

\date{Received: date / Accepted: date}

\maketitle

\begin{abstract}

Scaling-up from prototype systems to dense arrays of ions on chip, or vast networks of ions connected by photonic channels, will require developing entirely new technologies that combine miniaturized ion trapping systems with devices to capture, transmit and detect light, while refining how ions are confined and controlled. Building a cohesive ion system from such diverse parts involves many challenges, including navigating materials incompatibilities and undesired coupling between elements. Here, we review our recent efforts to create scalable ion systems incorporating unconventional materials such as graphene and indium tin oxide, integrating devices like optical fibers and mirrors, and exploring alternative ion loading and trapping techniques.

\keywords{ion traps \and quantum computation \and quantum information \and trapped ions \and ion-photon interface \and graphene \and indium tin oxide \and cavity cooling \and optical trapping \and micromirror \and motional heating \and CMOS ion trap \and hybrid trap \and scalable}

\end{abstract}

\section{Introduction: Creating systems}
\label{sec:Introduction: Creating systems}

\begin{figure}
\includegraphics[width=3.3in]{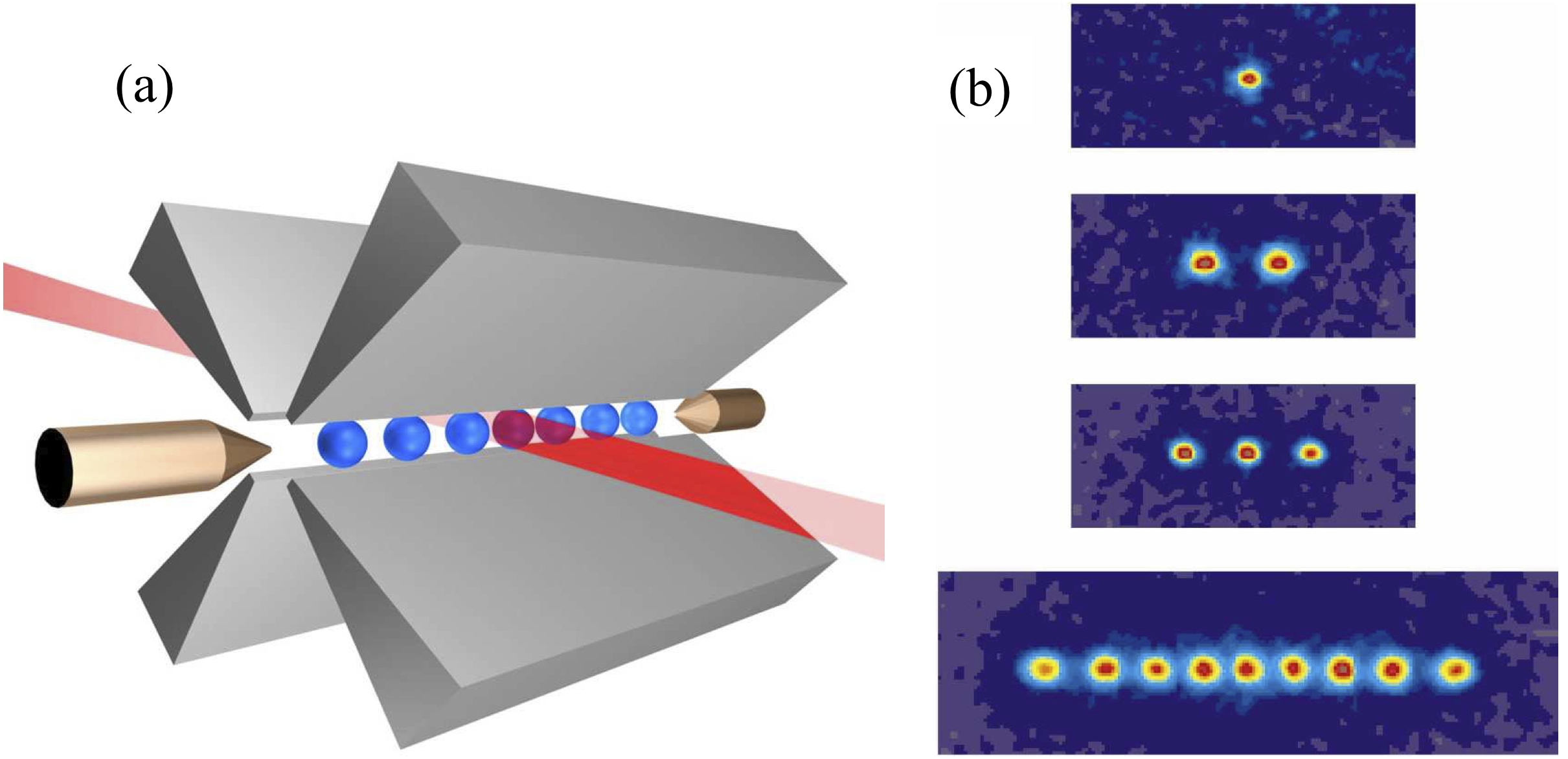}
\caption{The canonical four-rod Paul trap has been a workhorse for early demonstrations of QIP with ions. (a) Schematic of a Paul trap consisting of four radiofrequency (RF) electrodes and two end-cap (DC) electrodes to confine ions in a linear chain. A laser beam is shown applying a gate pulse to a single ion. (b) Camera images of few ions in a Paul trap. The spacing between adjacent ions is 2 to 5~$\mu$m. Images courtesy of University of Innsbruck.}
\label{fig:Innsbruck-trap}
\end{figure}

Originally conceived in the 1950s as a mass spectrometer, the Paul trap \cite{Paul1953} was reimagined as a platform for quantum information processing (QIP) in the 1990s \cite{Cirac1995,Monroe1995b}, when it was recognized that trapped atomic ions can be used as quantum bits (qubits). In pioneering experiments \cite{Wineland1998a}, a small qubit register was formed by a chain of ions confined within a Paul trap of precisely-machined metal rods or blades in ultrahigh vacuum (see Fig.~\ref{fig:Innsbruck-trap}); the ions' internal (electronic) and external (motional) states were manipulated using laser or microwave pulses to carry out quantum gates \cite{Nielsen2000,Mintert2001,Ospelkaus2008,Ospelkaus2011,Timoney2011}, which were then read-out via the fluorescence emitted by the ions. With this hardware, quantum protocols such as teleportation \cite{Barrett2004a,Riebe2007,Olmschenk2009b}, state mapping from an ion to a photon \cite{Stute2013}, and quantum simulation \cite{Kim2010c,Barreiro2011a,Lanyon2011,Gerritsma2011a,Britton2012a,Blatt2012,Islam2013} have been realized, and quantum algorithms including the quantum Fourier transform \cite{Chiaverini2005,Schindler2013}, the Deutsch-Josza algorithm \cite{Gulde2003}, Grover search \cite{Brickman2005}, and quantum error correction \cite{Schindler2011} have been demonstrated.

Trapped ions make for good qubits because of their long lifetimes (single ions can remain trapped for hours or days), long coherence times (ranging from ms to s) relative to the gate times (typically $\mu$s or less for single-qubit gates, and 10s~of~$\mu$s for two-qubit gates), strong inter-ion interactions (via the Coulomb force), and natural reproducibility. Most of the ingredients considered essential for quantum computation (expressed succinctly in DiVincenzo's criteria \cite{DiVincenzo1995,DiVincenzo2000}) have been realized with small numbers of ions, including: high-fidelity quantum state preparation \cite{Monroe1995,King1998,Roos1999}, single qubit rotations \cite{Roos1999,Nagerl1999}, multi-qubit operations \cite{Monroe1995a,Sackett2000,Leibfried2003,Schmidt-Kaler2003,Monz2009}, and state read-out \cite{Roos1999,Rowe2001,Myerson2008}. These proof-of-concept demonstrations underscore the potential for QIP with trapped ions, and as a result, have inspired intense technology development with the goal of refining and vastly scaling up trapped-ion systems (another milestone mentioned by DiVincenzo is a ``system [...] extendible to a large number of qubits" \cite{DiVincenzo2000}).

Designing a massive, versatile trapped-ion system that can retain coherence through a quantum computation necessitates exploiting new manufacturing platforms and devices, and understanding the microscopic behaviors of different trap materials. Hence, the challenge faced by researchers in recent years has been in crafting cohesive ion trap {\it systems}: combining diverse parts into a more powerful whole which achieves the ultimate performance from each component simultaneously.

One development in particular has become central to this research: the invention of the surface-electrode ion trap (or ``surface trap") \cite{Chiaverini2005a}, in which a single ion or chain of ions can be confined 20~$\mu$m to 2~mm above a trap chip \cite{Seidelin2006}. Surface traps can be constructed using standard printed circuit board manufacturing \cite{Brown2007}, or microfabrication techniques, making them highly versatile, and more reproducible than their precisely-machined, three-dimensional counterparts. As such, surface traps are more amenable to being shrunken in size and replicated to create an array of trapping sites. This compactness comes at the cost of shallower trap depths and greater anharmonicity of the trapping potential. Nonetheless, this convenient platform has allowed researchers to incorporate a variety of useful devices into ion traps.

Our work approaches scalable system design through a variety of means, almost all of which build on a surface trap tailored to a particular application. In attempting to control errors and noise, ion traps incorporating exotic materials such as graphene \cite{Eltony2014} and superconductors \cite{Wang2010} have been created. Towards scalable photonic interfaces, traps integrating devices including photodetectors \cite{Eltony2013}, optical fibers \cite{Kim2011c}, and high finesse cavities \cite{Cetina2013} have been designed and implemented. Through this trap integration, we seek to move beyond proof-of-principle experiments to machines encompassing thousands of manipulable atoms. But bringing these different components together presents a range of challenges from materials incompatibility to undesired noise. Many of our systems bridge atomic physics and other fields such as condensed matter physics or nanotechnology, often pushing the boundaries in more than one.

In this text, we review the many, often highly unconventional, ion systems we have developed, the challenges inherent in creating them, and new possibilities resulting from their advancement. We begin by discussing how materials science techniques applied to trap electrodes might be used to mitigate motional decoherence, then we describe the integration of a number of optical devices, including cavities, into ion traps for efficient light delivery to and from ions and coherent ion-photon coupling. Next, we turn our attention to the optical lattice as an alternate trap framework, and then to efforts to create hybrid systems combining neutral atoms and ions. Finally, we discuss work aiming to leverage a commercial CMOS (complementary metal-oxide-semiconductor) process to develop an integrated ion trap architecture.

\section{Mitigating motional decoherence}
\label{sec:Mitigating motional decoherence}

Miniaturization of traps is a natural first step towards realizing trapped-ion quantum computation at scale; however, such miniaturization faces the hurdle of increasing electric field noise as ions are brought closer to the electrode surface, resulting in motional state heating \cite{Turchette2000}. This is problematic because the shared motional state of ions in a trap forms a convenient ``bus" used for nearly all multi-qubit gates; if the motional state is altered by noise during a gate operation, this will limit the fidelity. Perplexingly, the heating rates observed in experiments are larger than expected from thermal (Johnson) noise from resistance in the trap electrodes or external circuitry \cite{Turchette2000,DeVoe2002}.

The origin of this excessive noise still eludes researchers, but the heating rate has been observed to diminish by about $100$ times when trap electrodes are cooled from room temperature (295~K) to 4~K, suggesting that it is thermally activated \cite{Deslauriers2006,Labaziewicz2008,Labaziewicz2008b}. The data are also consistent with a $d^{-4}$ scaling law (where $d$ is the distance from the ion to the nearest electrode), as would be predicted for a random distribution of fluctuating charges or dipolar patch potentials on the surface (for Johnson noise, a $d^{-2}$ scaling law would be expected) \cite{Deslauriers2004,Dubessy2009,Safavi-Naini2011}. Theoretically, a correlation length associated with surface disorder has been shown to characterize electric field noise generated near the surface \cite{Dubessy2009,Low2011} and microscopic models have been developed based on fluctuations of the electric dipoles of adsorbed molecules \cite{Daniilidis2012,Safavi-Naini2011}. However, the predictions of available noise models are not in agreement with recent measurements \cite{Bruzewicz2015}. Experimentally, argon-ion-beam cleaning of the trap surface in-vacuum resulted in an astounding 100-fold reduction in the heating rate \cite{Hite2012,Hite2013,Daniilidis2014}, suggesting that surface effects play a primary role.

In this section, we review efforts to attain a deeper understanding of this phenomenon, through studies of motional heating in traps composed of qualitatively different materials, such as graphene. These efforts complement work (described above) focused on altering the electrode composition in situ \cite{Hite2012,Hite2013,Daniilidis2014}. By bringing ions close to the surfaces of exotic materials, and fabricating traps from materials/technologies used for solid-state qubits, these investigations also make progress towards new interfaces between ions and condensed matter systems, a first step towards a new hybrid quantum system.

\subsection{Superconducting ion traps}
\label{sec:Superconducting ion traps}

The challenge posed by excessive motional heating motivated Wang et al.~\cite{Wang2010} to develop ion traps incorporating superconducting materials. They reasoned that because a superconductor expels electric fields, noise sources within superconducting electrodes would be screened from the ion, while noise sources on the surface would be unaffected. In their measurements with superconducting traps, described here, they found no significant change in the heating rate as electrodes crossed the superconducting transition, suggesting that motional heating is predominantly independent of noise sources in the bulk of a material.

Initially, it was unclear whether crafting such a system would be possible. Since the electrodes would be superconducting thin films, light scattered from short wavelength lasers needed for ionization, cooling and state detection could create quasiparticles in the electrodes, driving them into the normal state. Further, the critical current of the superconducting electrodes had to be large enough to carry the RF trap drive current.

\begin{figure}
\includegraphics[width=3.3in]{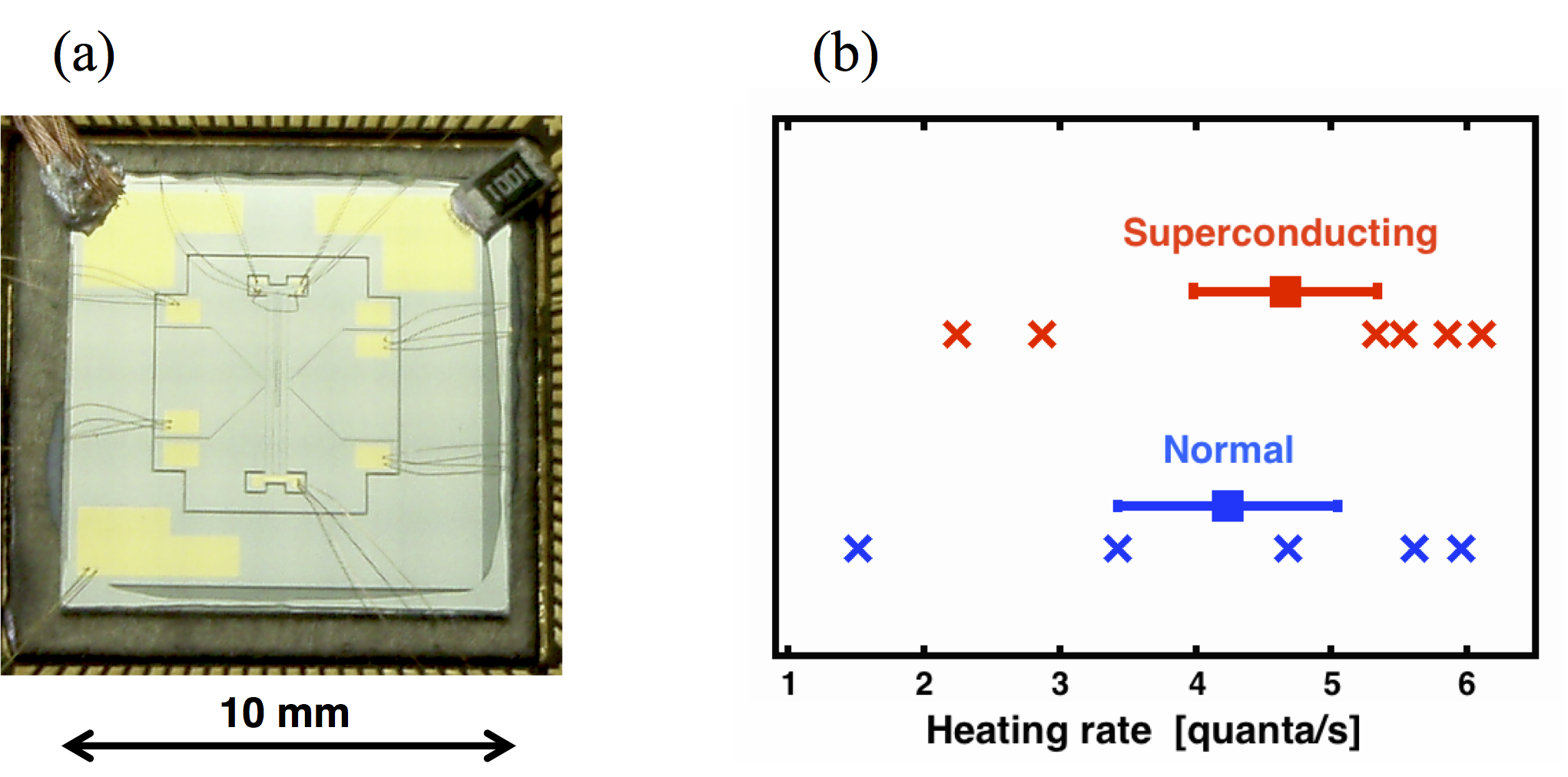}
\caption{An ion trap with superconducting electrodes has been produced and characterized, revealing that the heating rate is unchanged across the superconducting transition. Adapted from Ref.~\cite{Wang2010}. (a) Photograph showing one of the niobium ion traps tested, mounted in a ceramic pin grid array (CPGA) carrier. A single ion is confined 100~$\mu$m above the center of the trap. A heating resistor can be seen in the top right corner, and a copper braid to thermally anchor the trap can be seen in the top left corner. The gold contact pads are added for wirebonding. (b) Plot showing heating rate measurements for the pictured trap in the normal and superconducting states. Data were taken at temperatures within $\sim2$~K of the superconducting transition temperature. The crosses are different measurements taken over the course of 2 days: blue data points were taken with electrodes in the normal state, and red data points were taken with electrodes in the superconducting state. Filled squares show the mean heating rate measured (with error bars at 1 standard deviation). The difference between normal and superconducting data is not significant.}
\label{fig:Supercon-figure}
\end{figure}

With ion traps fabricated from 400~nm of niobium or niobium nitride on a sapphire substrate, Wang et al.~\cite{Wang2010} were able to realize stable, single-ion confinement and to confirm the electrodes were in fact superconducting \cite{Wang2010}. By sweeping the trap temperature, they compared heating rates measured with normal versus superconducting electrodes (the transition temperature was estimated to be near to 10~K). Overall, the heating rates observed were comparable to previous measurements with normal metal traps (Au, Ag, Al electrodes) \cite{Labaziewicz2008,Labaziewicz2008b} and did not change significantly across the superconducting transition (see Fig.~\ref{fig:Supercon-figure}).

These results suggest that motional heating is unrelated to bulk resistivity. The London penetration depth (which is about an order-of-magnitude less than the 400~nm film thickness for Nb) defines what is meant by surface versus bulk. In this sense, it seems that motional heating is predominantly a surface effect, and is largely independent of properties of the bulk of the electrodes. These results have been corroborated by Chiaverini et al. \cite{Chiaverini2014} in a series of experiments looking at the heating rates in niobium and gold traps over a temperature range spanning about 4~K to room temperature (295 K). They found no discernible difference in motional heating between the two types of trap over this temperature range.

In addition to providing information about the mechanism of motional heating, this investigation also established the feasibility of combining ion systems with superconducting materials, an important step towards creating interfaces between trapped ions and superconducting devices. If a coherent coupling between a superconducting qubit (such as a flux qubit) \cite{Devoret2013} and a trapped-ion qubit, embodied in a microwave energy hyperfine spin state, can be demonstrated, this would allow for the development of hybrid quantum systems able to capitalize on the strengths of both types of qubit \cite{Tian2004,Kielpinski2012}. Similarly, superconducting qubits could potentially be used to realize an all-electrical interface to the rotational states of molecular ions \cite{Schuster2011}. The ability to shuttle a trapped ion across a planar trap also suggests the possibility to use a trapped ion as a ``read/write head" for another, perhaps less accessible, superconducting quantum device \cite{Cirac2000,Hensinger2006}.

\subsection{Ion traps coated with graphene}
\label{sec:Ion traps coated with graphene}

The study with superconducting ion traps revealed that surface effects play a significant role in motional heating, inspiring Eltony et al.~\cite{Eltony2014} to seek other materials with unique surface properties. Graphene was a material of interest because it has an exterior free of dangling bonds and surface charge \cite{Wang2008}. Graphene also has the ability to passivate metals from oxidation and other surface adsorbates \cite{Sutter2010,Chen2011}. To investigate, Eltony et al.~\cite{Eltony2014} produced an ion trap coated with graphene, and measured motional heating in the trap. Surprisingly, they discovered that despite the relatively inert surface, the heating rate in the graphene-coated trap was significantly worsened relative to a comparable, uncoated trap, which may be indicative of the noise source.

\begin{figure}
\includegraphics[width=3.0in]{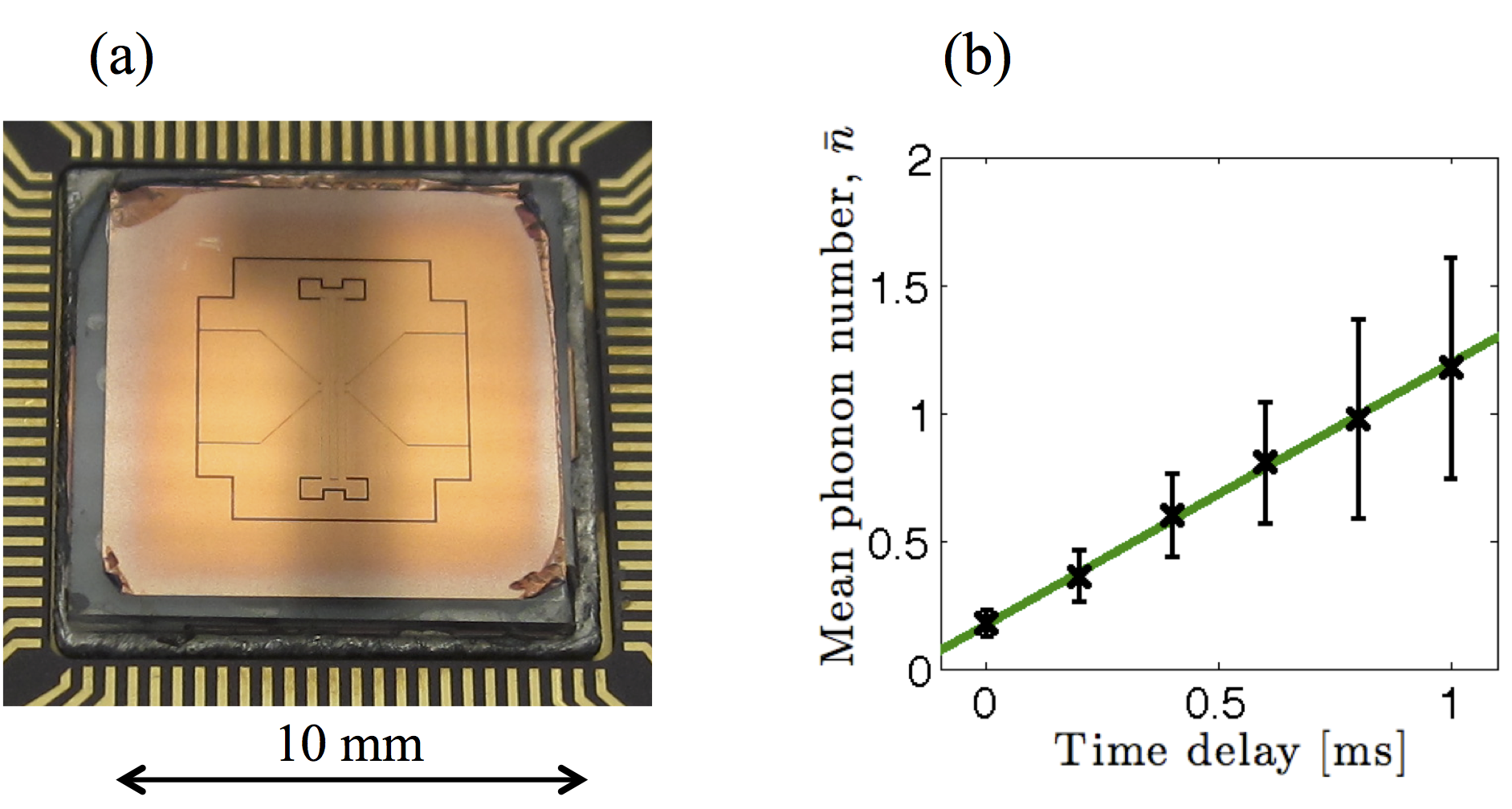}
\caption{A method to cover copper trap electrodes with graphene was developed, and motional heating was studied in such a trap, yielding the unexpected result of significantly worsened heating (relative to a similar, uncoated trap). Adapted from Ref.~\cite{Eltony2014}. (a) Photograph showing one of the graphene-coated ion traps tested, mounted in a CPGA. A single ion is confined 100~$\mu$m above the center of the ground electrode adjacent to the notch in the RF electrode. (b) Plot showing the inferred mean vibrational quantum number as a function of time (for the the lowest frequency mode of the ion at 0.86~MHz). The linear fit indicates a heating rate of $1020\pm30$ quanta/s, which is about 100 times higher than expected for a bare metal trap operated under comparable conditions.}
\label{fig:Graphene-figure}
\end{figure}

Integrating graphene into an ion trap posed some fabrication difficulties: the trap electrodes would require sufficient electrical conductivity to support the RF trap drive without damaging graphene, and a typical surface trap has dimensions on the order of 1~cm$\times$1~cm, which precluded using the standard peel-off method \cite{Novoselov2004} to produce graphene. To meet these challenges, Eltony et al.~\cite{Eltony2014} devised a procedure to synthesize graphene directly onto a pre-patterned copper trap, which acted as the seed metal for chemical vapor deposition (CVD). The copper electrodes beneath graphene also served to carry the radio frequency trap drive current (see Fig.~\ref{fig:Graphene-figure}). This process included pre-cleaning and annealing steps to remove potential contamination or copper oxide on the trap surface before graphene growth. It should be noted that CVD synthesis leaves the underlying metal electrodes with an increased surface roughness (RMS roughness increased by about an order of magnitude in this process). This roughening could be avoided by transferring graphene from a separate growth substrate onto the trap \cite{Li2009,Kim2009}.

Large graphene films grown on metal using CVD techniques similar to those Eltony et al.~\cite{Eltony2014} employed have been shown to protect the underlying metal surface from oxidation and reaction with air, even following heating \cite{Sutter2010,Chen2011}. Thus, they hypothesized that the graphene-coated ion trap would exhibit lower electric field noise, and hence induce a lower heating rate. In practice, quite the opposite occurred: the average heating rate measured was $\approx$100 times larger than expected for a bare metal trap operated under comparable conditions! \cite{Labaziewicz2008,Labaziewicz2008b,Eltony2014}

It is possible that the graphene-coating failed to reduce heating because the particular surface contaminants it prevents, such as metal oxide, and carbon or oxygen adsorbates, are not important noise contributors. But why was motional heating {\it exacerbated} in the graphene-coated trap? Other investigations have raised suspicion that hydrocarbon molecules adsorbed onto the trap surface during the initial vacuum bake are important contributors to motional heating \cite{Hite2012,Hite2013,Daniilidis2014}. Perhaps hydrocarbon contamination, which is ubiquitous on graphene, contributed significantly \cite{Meyer2008}.

Overall, it appears that adding a graphene coating is unlikely to reduce anomalous heating, but it may prove fruitful for further studies by affording some control over the type of surface contaminants present on the trap. Hydrocarbon content on graphene can be monitored via atomic-resolution, high-angle dark field imaging and electron-energy-loss spectroscopy \cite{Bangert2009}. Because graphene coating prevents reaction of underlying metal electrodes with environmental contaminants (other than hydrocarbon residues) \cite{Sutter2010,Chen2011}, it may be a good material choice for further studies investigating motional heating in traps with different types and arrangements of hydrocarbon deposits on the surface.

\section{Incorporating optical components}
\label{sec:Incorporating optical components}

Efficient and scalable interfacing between ions and photons is essential for a variety of reasons, notably: 1) the ion-light interface typically provides most of the state preparation, control and readout in a trapped-ion quantum processor, and 2) efficient coupling between ions and photons can pave the way for scalable quantum processors and networks consisting of trapped-ion quantum registers (stationary qubits) with photonic interconnects (``flying" qubits) \cite{Kimble2008}. In terms of state preparation, laser light is used for motional cooling and internal state initialization. In terms of control, it is used for single-qubit and multi-qubit operations. In terms of readout, ion fluorescence needs to be detected for quantum state measurement and for probabilistic entanglement \cite{Luo2009,Maunz2009}, and the speeds of these processes are limited by the rate of fluorescence detection. This warrants an ion trapping infrastructure that provides good optical access, individual optical addressability of ions, efficient delivery and collection of light from different ions, and finally, compactness and scalability of the optical elements to large ion numbers. Meanwhile, a coherent ion-photon interface, necessary for deterministic entanglement across photonic interconnects, has much more stringent requirements of photon confinement to small spatial and spectral regions. 

One frontier for addressing all these requirements and for scaling up QIP is the integration of optical elements for light delivery, collection and photon confinement with the ion traps in vacuum. Two major challenges come with this integration: alignment of optical elements  with tiny modes to point emitters, and trapping charged particles close to dielectric surfaces. Integrated optics are not easily made reconfigurable, so their alignment to the ions must be partially built into the trapping architecture. The second problem is that dielectric surfaces are susceptible to light-induced charging, which results in strong and difficult-to-control forces on the ions, inducing micromotion, large displacements or even making the ions untrappable \cite{Harlander2010,Wang2011,Sterk2012}. Nonetheless, these challenges have started to be addressed in the last several years by a number of groups integrating various optical elements with ion traps, including microfabricated phase Fresnel lenses \cite{Jechow2011,Streed2011}, embedded micromirrors \cite{Herskind2011,TrueMerrill2011} and fibers \cite{VanDevender2010,Kim2011c}, transparent trap electrodes \cite{Eltony2013}, nanophotonic dielectric waveguides \cite{Mehta2015}, macroscopic optical cavities \cite{Guthohrlein2001,Mundt2002,Stute2012b,Herskind2009a,Cetina2013}, and microscopic, fiber-based cavities \cite{Steiner2013}.

The efficiency of photon coupling to an ion tends to improve with smaller distance $d$ from the ion to the optical element as $d^{-2}$ in the far-field, given that the maximal dimensions of the optical element are constrained. In the case of light collection, the collection efficiency is given by $\Omega/(4\pi)\propto d^{-2}$, where $\Omega$ is the solid angle subtended by the collection optic. In the case of light delivery, the absorption probability is approximately the ratio of the resonant atomic photon scattering cross-section $\frac{3}{2\pi}\lambda_0^2$ (where $\lambda_0$ is the transition wavelength) to the beam area $\frac{1}{2}\pi w^2$. This ratio is usually much smaller than unity, but can be maximized by reducing the mode waist $w\propto d$. To achieve coherent coupling, a cavity of finesse $F$ can be used to enhance the effective interaction cross-section by $F/\pi$, the number of round trips a photon makes in this cavity, resulting in the cooperativity $\eta \equiv 4g^2/(\kappa\Gamma) = 2\cdot(F/\pi)\cdot(\frac{3}{2\pi}\lambda_0^2)/(\frac{1}{2}\pi w^2)$ for a single two-level atom at the antinode of the cavity mode  \cite{HarukaAdvances2011}. The cooperativity $\eta$ is the ratio of the coherent process rate, characterized by the single photon Rabi frequency $2g$, to the rates of incoherent processes: the atomic spontaneous emission rate $\Gamma$ and the cavity decay rate $\kappa$. To reach the strong coherent coupling regime of $\eta\gg 1$, once again small cavity mode waist $w$ is desired, which is proportional to $d^{1/2}$ for a fixed cavity type \cite{Siegman1986}. Thus, in many cases of light collection, delivery and coherent coupling, the requirement of small ion distance to dielectric optical elements has to be reconciled with the issues related to light-induced charging of dielectrics.

In this section, we review four systems integrating optics and trapped ions while addressing some concerns with proximity of dielectrics and with alignment of small-waist modes of light to trapped ions: a planar trap embedded with an optical fiber, a planar trap with transparent electrodes, a planar trap fabricated on a high-reflectivity mirror, and a planar trap confining long ion chains within a macroscopic optical cavity.

\subsection{Laser delivery via integrated optical fiber}
\label{sec:Laser delivery via integrated optical fiber}

When we think about collecting and delivering light efficiently to a space with limited optical access, optical fibers, which offer low loss in a compact package, immediately come to mind. But there are challenges to be overcome in putting a fiber into an ion trap: transporting light efficiently to or from an ion requires accurate spatial overlap of the ion's optical cross-section with the field mode of the fiber. Additionally, dielectric materials comprising the fiber may result in charging near the fiber tip (particularly as lasers pass through) which distorts the trapping potential \cite{Harlander2010,Wang2011}. VanDevender et al. \cite{VanDevender2010} were first to demonstrate an embedded optical fiber; they collected fluorescence from an ion trapped above the surface through an integrated multi-mode fiber. Brady et al. \cite{Brady2011} created an ion trap chip incorporating an array of optical fibers and confirmed ion trapping and preliminary light detection with their system, establishing a proof-of-concept for a large fiber array trap architecture. Following this work on light collection, Kim et al.~\cite{Kim2011c} demonstrated light {\it delivery} through an integrated single-mode (SM) fiber.

\begin{figure}
\includegraphics[width=3.3in]{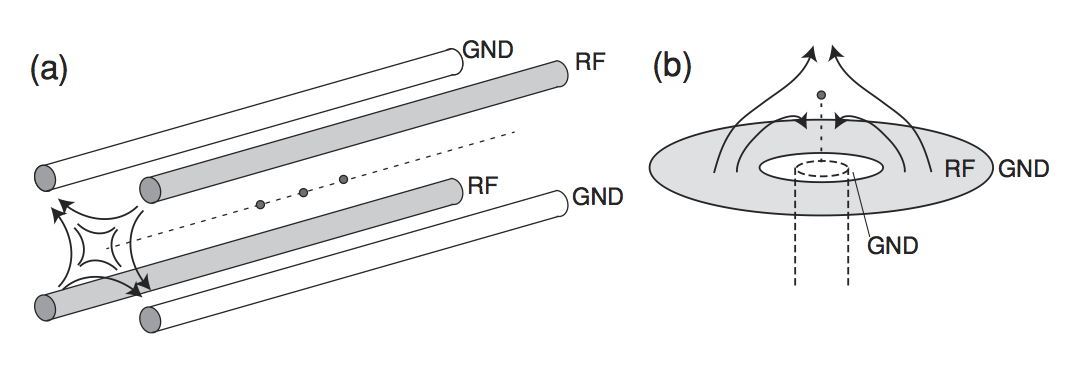}
\caption{This illustration compares the traditional four-rod linear Paul trap (a) to the point Paul trap (b). The latter achieves quadrupole ion confinement through RF on a single, ring-shaped electrode. Dashed lines suggest how cylindrical elements, such as optical fibers, may be introduced to the point Paul geometry. Reproduced from Ref.~\cite{Kim2010b}.}
\label{fig:Ring-trap}
\end{figure}

To accommodate an optical fiber, Kim et al.~\cite{Kim2011c} chose to deviate from the standard, linear Paul trap layout and instead to develop a trap in which the quadrupole potential is generated by a planar RF ring (a concept first explored analytically by Brewer et al. in 1992 \cite{Brewer1992}). The axial symmetry of this kind of trap is naturally suited to optical fiber integration in that the fiber can be placed at the center of the trap (the symmetry point) without breaking the symmetry of the trapping fields. Another benefit is that the RF ring alone provides confinement (unlike in a linear trap, where DC end caps are also required), a simplification which also means that RF translation can be performed without the added complexity of solving for appropriate DC trap voltages at each point in the trajectory. However, this design does not allow for micromotion-free DC translation along one trap axis, as a linear trap does. Kim et al.~\cite{Kim2010b} modeled and verified the circular surface trap design shown in Fig.~\ref{fig:Ring-trap} experimentally prior to attempting fiber integration.

\begin{figure}
\includegraphics[width=2.8in]{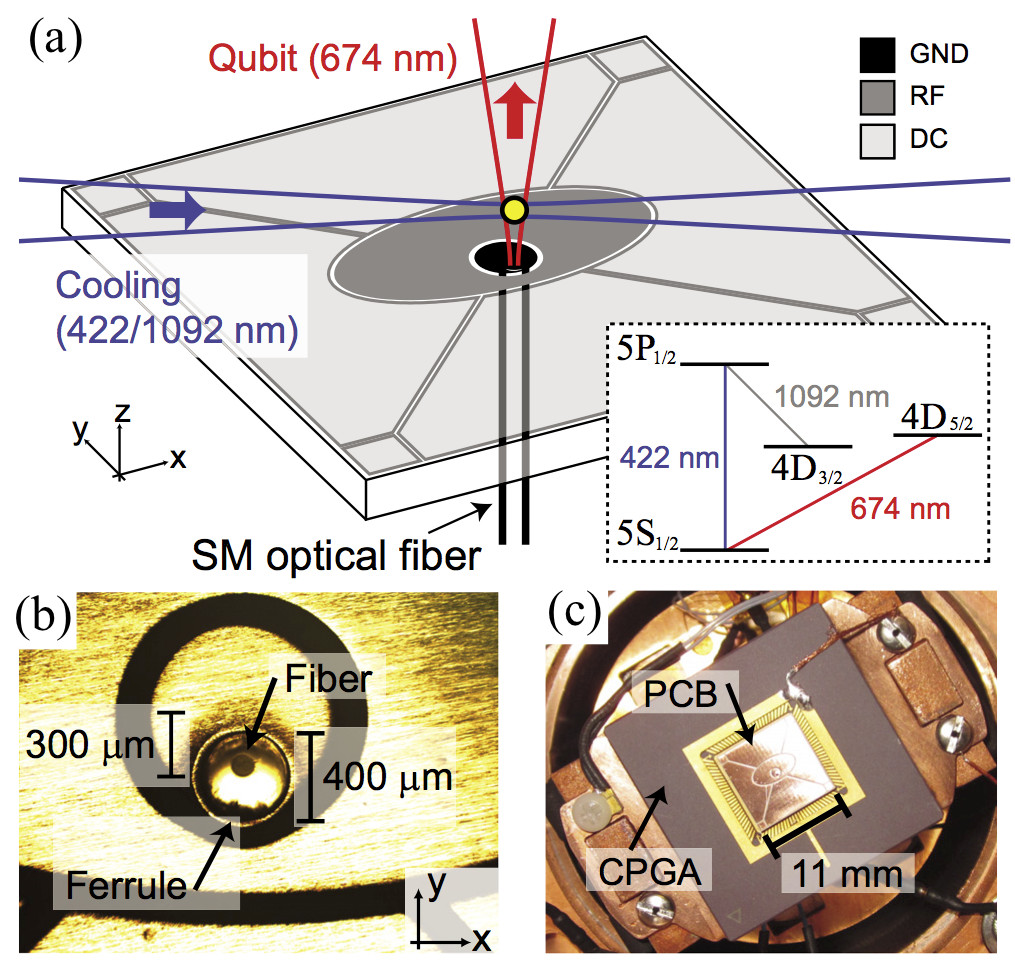}
\caption{A single-mode optical fiber embedded into a surface trap has been used to deliver light to a single ion, successfully addressing the qubit transition. Reproduced from Ref.~\cite{Kim2011c}. (a) Schematic of the surface-electrode ion trap with integrated optical fiber. The $^{88}$Sr$^+$ qubit laser (shown in red) is delivered through the fiber, while Doppler cooling beams (shown in blue) propagate in free space, parallel to the trap surface. Inset: the relevant level structure for $^{88}$Sr$^+$. (b) Image showing how the optical ferrule is oriented with respect to the trap electrodes. The fiber is aligned with the minor-axis of the trap to match the ion position. (c) Photograph showing the fiber-trap system mounted in a CPGA and installed in the experimental vacuum chamber.}
\label{fig:Fiber-figure}
\end{figure}

For stable fiber-coupling, the ion must remain as stationary as possible, meaning that micromotion should be minimized \cite{Berkeland1998}. So not only should the ion be aligned with the optical fiber, it must also be positioned so that it sits close to the RF null of the trapping field. To achieve this two-fold alignment, Kim et al.~\cite{Kim2011c} repositioned the ion through RF translation (thereby avoiding micromotion), by adjusting an RF voltage in phase with the trap RF applied to a neighboring triangular electrode (see Fig.~\ref{fig:Fiber-figure}) \cite{Herskind2009,VanDevender2010,Kim2010b,Kim2011c}. Lastly, they created an asymmetry in the trap layout by choosing ellipsoidal rather than circular RF electrodes, which simplified laser cooling by tilting the principal axes of the trap so that a single laser beam parallel to the surface had a projection onto all three axes.

\begin{figure}
\includegraphics[width=3.0in]{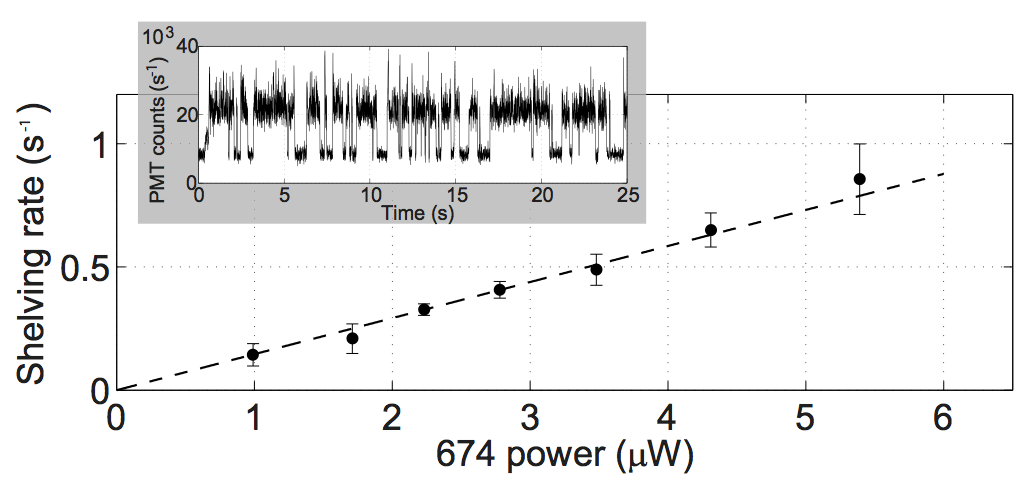}
\caption{A single-mode optical fiber embedded into a surface trap has been used to deliver light to a single ion, successfully addressing the qubit transition (5S$_{1/2}$~$\leftrightarrow$~4D$_{5/2}$). This plot shows the shelving rate into the 4D$_{5/2}$ state as a function of 674~nm (qubit) laser power coupled into the trap fiber. Inset: telegraph log of a single ion as it is shelved into the dark 4D$_{5/2}$ state by 5.4~$\mu$W of fiber light. Reproduced from Ref.~\cite{Kim2011c}.}
\label{fig:Fiber-data}
\end{figure}

Kim et al.~\cite{Kim2011c} were ultimately able to deliver laser light to a single ion confined 670~$\mu$m above the trap surface, effectively addressing the qubit transition (see Fig.~\ref{fig:Fiber-data}). This configuration is compatible with photonic crystal fibers which could be used to transmit all required lasers through a single integrated port. Using a lensed fiber tip to increase the effective numerical aperture (NA) and adding antireflection coatings to optical surfaces could significantly improve the net system collection efficiency so that it is comparable to the best conventional collection schemes, which employ a high numerical aperture
objective near the ion, requiring extensive solid angle access, plus a photomultiplier or charge-coupled detector located outside of the vacuum chamber.

Integrating multiple fibers into a surface trap is a promising means to simultaneously address, cool, and readout ions within a multi-ion architecture, particularly within miniaturized systems where optical access with conventional means is prohibitively difficult. Integrated fibers are also being explored for use in quantum networks \cite{Stute2012a,Stute2012b,Takahashi2013}, where they can facilitate matching of the spectral and spatial modes of the flying qubit (photon) to the time-inverted emission properties of the memory qubit (atom). In a similar vein, the small radii possible with fiber mirrors, which allows for short confocal cavities with a small mode area (or tight waist) and hence stronger ion-cavity coupling, has motivated work creating integrated optical cavities with fibers \cite{Colombe2007,Steiner2013,Brandstatter2013} (see Section \ref{sec:Ion traps with optical cavities}).

\subsection{Transparent traps}
\label{sec:Transparent traps}

Efficient fluorescence collection is essential for fast, high-fidelity state detection and ion-photon entanglement \cite{Luo2009}. Because the emission pattern of a trapped-ion qubit is isotropic under most conditions, this means capturing the largest solid angle possible. Yet in a surface trap, the $2\pi$ solid angle obstructed by the trap is typically inaccessible (with the exception of trap designs incorporating a micromirror embedded into the trap beneath the ion \cite{Herskind2011,TrueMerrill2011,Noek2010}, see Sec.~\ref{sec:Embedded mirrors}). To circumvent this, Eltony et al.~\cite{Eltony2013} created a transparent surface trap and collected ion fluorescence {\it through} it.

Transparent electrodes employed commercially in touch screens and LCD displays are typically made of indium tin oxide (ITO), a material that is both an electrical and an optical conductor. However, using ITO electrodes in an ion trap poses some difficulty because ITO has a resistivity about 1000 times higher than metals typically used for trap electrodes (at room temperature), and ITO is an oxide, making laser-induced charging of its surface a concern \cite{Harlander2010,Wang2011}. To construct a workable transparent ion trap, Eltony et al.~\cite{Eltony2013} sputtered 400~nm of ITO for all trap electrodes, and then added 5~nm of evaporated gold on the RF electrodes to enhance their conductivity nearly 1000 fold while still maintaining about $60\%$ optical transparency at the fluorescence wavelength.

Eltony et al.~\cite{Eltony2013} verified stable trapping of single ions, and did not observe worsened charging of the ITO trap, even after a deliberate attempt to charge up the center of the trap with a 405~nm laser beam of $\approx5$~mW/mm$^2$ intensity.

\begin{figure}
\includegraphics[width=2.7in]{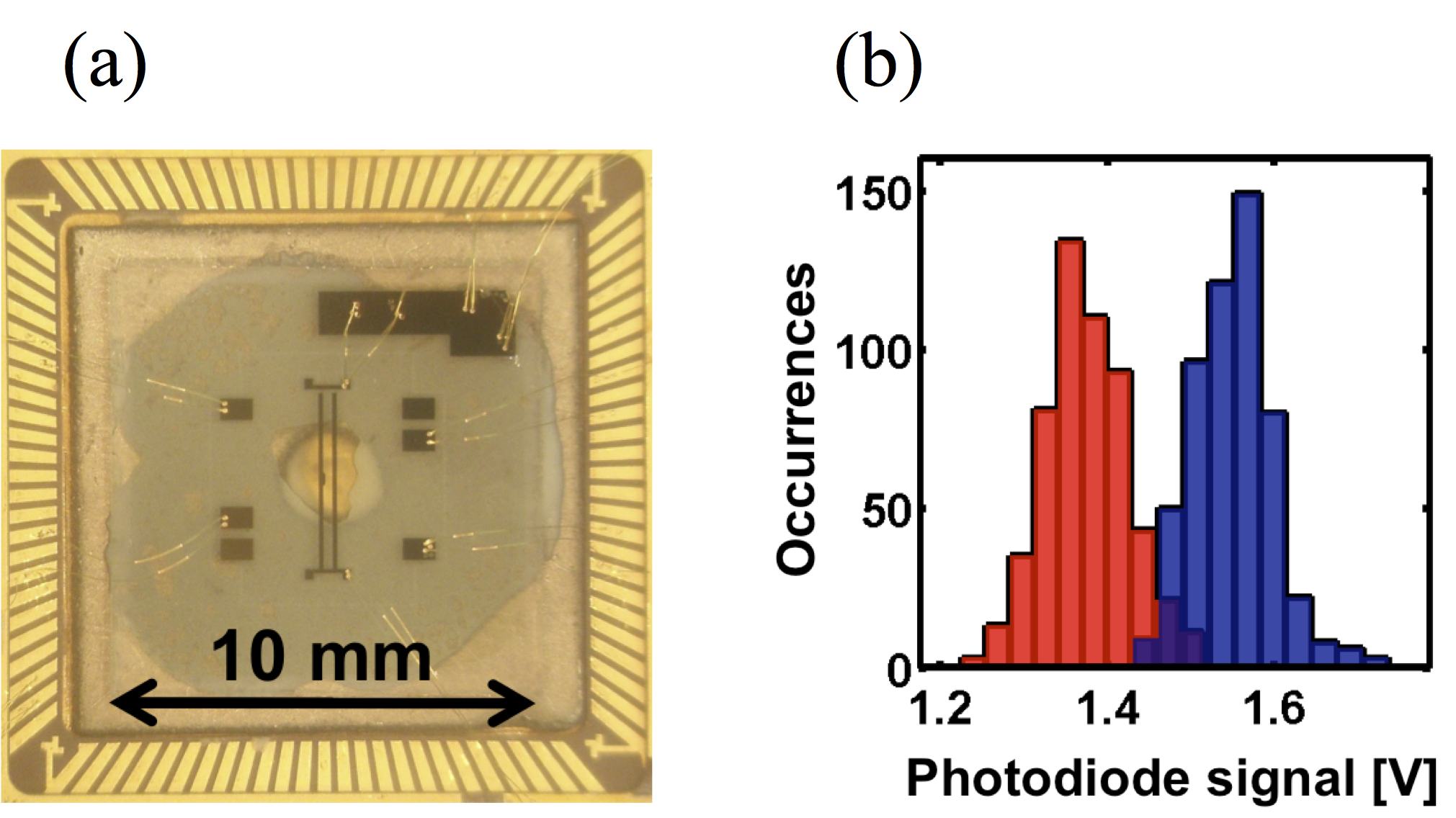}
\caption{The first transparent ion trap was developed, and fluorescence from a cloud of ions confined above was transmitted through the electrodes and detected using a photodiode sandwiched below the trap. Adapted from Ref.~\cite{Eltony2013}. (a) An ITO trap mounted in a CPGA; 50~nm of Au (5~nm was used in a later version) is visible on the RF electrodes and contact pads for wire bonding. A metal spacer mounted below (with epoxy) is visible through the trap. (b) Histogram of photodiode voltages over a period of several minutes without ions (red), and after loading an ion cloud (blue). The ion cloud fluorescence is distinguishable from the background.}
\label{fig:ITO-figure}
\end{figure}

As a proof-of-concept demonstration of light collection, Eltony et al.~\cite{Eltony2013} sandwiched a transparent ion trap above a commercially available PIN photodiode, which was then used to detect fluorescence from a cloud of $\approx50$ ions (see Fig.~\ref{fig:ITO-figure}) confined above. In principle, nearly 50\%~ of the total solid angle could be captured by using a transparent trap and a photodiode with a large active area (or by focusing fluorescence onto the photodiode using additional optics below the trap). For commercially available ITO films, optical losses can be as low as 10\% \cite{Kim1999}. Combining this with a photodetector such as the Visible Light Photon Counter, which has a quantum efficiency of 88\% at 694~nm and 4~K \cite{McKay2009}, could boost the total detection efficiency to nearly 40\%, far superior to the usual 1-5\% possible with conventional bulk optics and a photomultiplier placed outside the vacuum chamber. In this experiment, it would mean reducing the time for quantum state detection with 99\% fidelity from 200~$\mu$s to just 5~$\mu$s.

\begin{figure}
\includegraphics[width=2.2in]{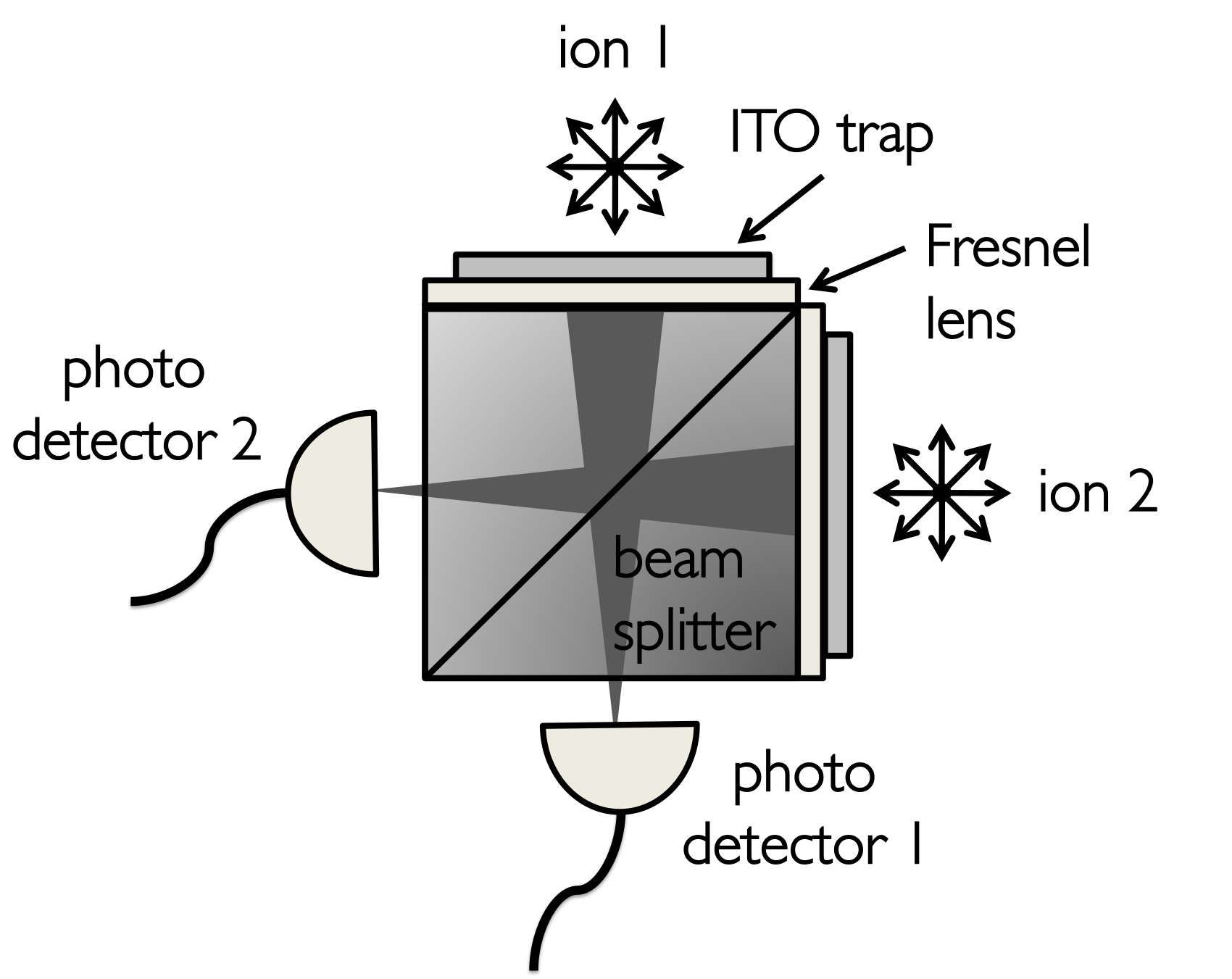}
\caption{An entanglement unit, pictured here, has been proposed based on a pair of transparent ion traps (see text). Reproduced from Ref.~\cite{Eltony2013}.}
\label{fig:Entanglement-unit}
\end{figure}

More generally, the ability to build transparent ion traps opens up many possibilities to incorporate devices to efficiently transfer and detect light from ions. To illustrate this potential, Eltony et al.~\cite{Eltony2013} proposed a compact entanglement unit (see Fig.~\ref{fig:Entanglement-unit}), in which transparent traps mounted on adjacent faces of a beam splitter house two (or more) ions to be entangled. Diffractive optics below the traps overlap the images of the two ions on detectors at opposite faces of the cube, allowing for heralded entanglement generation between the ions \cite{Maunz2009,Luo2009}. In theory (neglecting losses due to reflection at interfaces, and absorption in materials other than ITO), the coupling efficiency for photons in this proposed entanglement unit would be $\approx10$ times higher than in the current state-of-the-art \cite{Hucul2014}, resulting in $\approx100$ times the entanglement rate (since two photons need to be collected simultaneously).

\subsection{Embedded mirrors}
\label{sec:Embedded mirrors}

Similar to the transparent trap described above (Sec.~\ref{sec:Transparent traps}), traps integrating mirrors can make photon collection more efficient by expanding the solid angle captured. This integration has been realized in a number of ways using metallic mirrors \cite{Shu2009,Noek2010,Shu2011,TrueMerrill2011,Maiwald2012}. Dielectric mirrors, which offer higher reflectivity, could also be integrated to further enhance light collection, but more importantly, to build a small waist optical cavity around the ion. To achieve strong coherent coupling, low loss mirrors and a small cavity mode waist are required, which generally necessitates working with a short cavity formed by dielectric mirrors (see the introduction to Section~\ref{sec:Incorporating optical components}). However, short cavities, which place the ion very close to the mirrors have been avoided because of concerns about charge build-up on the dielectric mirror surface \cite{Harlander2010,Wang2011}. This can be circumvented via collective coupling of multiple ions to a longer cavity \cite{Herskind2009a}, at the cost of greater experimental complexity (see Section~\ref{sec:Ion traps with optical cavities}). To evaluate the actual impact of charging and achievable mirror finesse with a single ion confined close to a dielectric mirror, Herskind et al. \cite{Herskind2011} created a surface trap with an embedded, high-reflectivity mirror.

\begin{figure}
\includegraphics[width=3.3in]{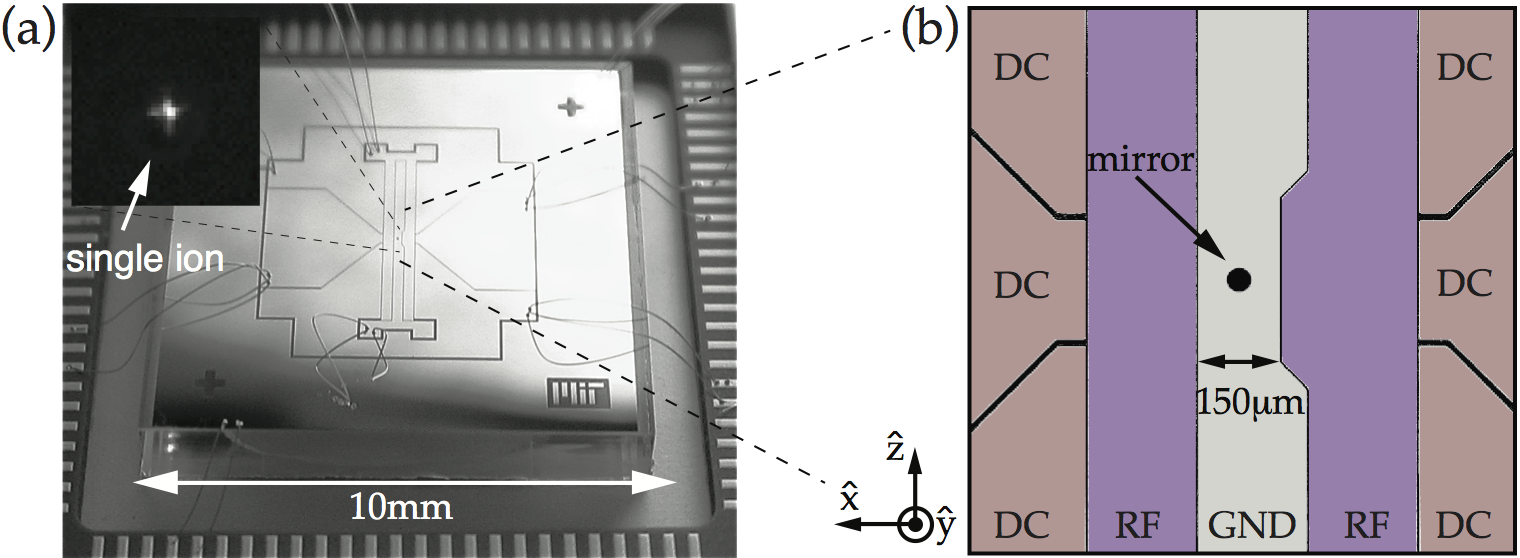}
\caption{An ion trap has been fabricated on a high-reflectivity mirror. Reproduced from Ref.~\cite{Herskind2011}. (a) Photograph showing the finished trap mounted in a CPGA. Inset shows a single trapped $^{88}$Sr$^+$ ion. (b) False color microscope image of the central region of the trap with electrodes labelled. The black circle shown is the mirror aperture, which has a diameter of 50~$\mu$m. An ion is trapped 169~$\mu$m above the mirror aperture.}
\label{fig:MirrorTrap}
\end{figure}

Silver electrodes were fabricated directly onto the surface of a planar dielectric mirror, which served as the trap substrate. To minimize dielectric exposure, the mirror surface was only exposed through a 50~$\mu$m - diameter circular aperture in the center of the trap (see Fig.~\ref{fig:MirrorTrap}). The mirror initially had a transmission coefficient of  45~ppm and scattering and absorption losses of 25~ppm at 422~nm (the wavelength at which the dielectric coating was optimized). Following trap fabrication, the additional optical losses incurred averaged 130~ppm, suggesting that the mirror quality was not significantly compromised by the process.

\begin{figure}
\includegraphics[width=3.3in]{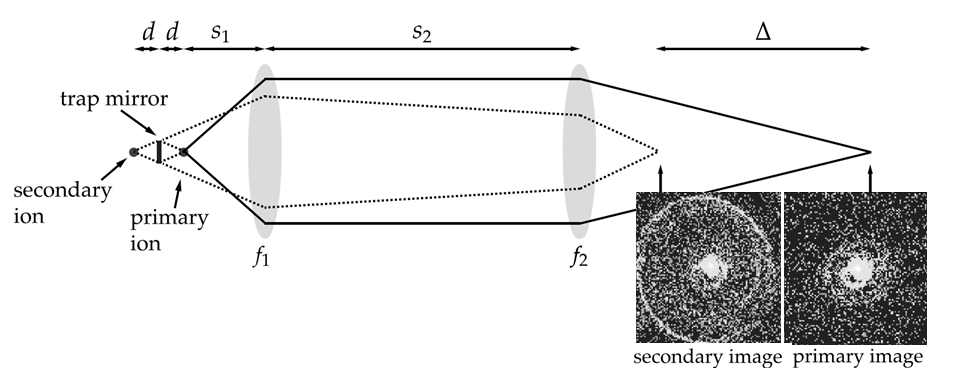}
\caption{A single ion and its mirror image have been observed in a trap fabricated on a high-reflectivity mirror. This schematic shows the trapped ion imaged directly (primary) and through the integrated mirror (secondary). Reproduced from Ref.~\cite{Herskind2011}.}
\label{fig:MirrorTrapImage}
\end{figure}

Herskind et al. \cite{Herskind2011} demonstrated stable trapping of single ions confined 169~$\mu$m above the mirror aperture, with Doppler-cooled ion lifetimes of several hours. Despite the proximity of the ion to the mirror surface, low sensitivity to laser-induced charging was observed for test wavelengths of 405~nm, 461~nm, and 674~nm. The strongest charging effect was measured following continuous illumination with a 405~nm laser (with power about 200~$\mu$W, focused to a 50~$\mu$m spot size) grazing the trap for 10 minutes, after which the induced electric field at the ion location was inferred to be 20~V/m. This field strength is comparable to that seen in metal traps without dielectric mirrors ~\cite{Wang2011}, suggesting that the close presence of dielectric mirrors would not significantly affect trapping in planar trap configurations. Basic functionality of the embedded mirror was confirmed in situ through observations of the ion and its mirror image at different focal points of the imaging system (see Fig.~\ref{fig:MirrorTrapImage}).

This system could be extended to form a cavity-ion system by adding a concave mirror above the trap. A concave mirror with a small radius of curvature could be inserted into the trap substrate, or into a second substrate above the trap, using laser machining \cite{Hunger2010} or chemical etching \cite{Noek2010,TrueMerrill2011}. The absence of adverse effects observed with exposed dielectric 169~$\mu$m away from the ion indicates that a sub-millimeter cavity length could be achievable, putting strong coupling of a single ion within the realm of possibility.

\subsection{Ion traps with optical cavities}
\label{sec:Ion traps with optical cavities}

While strong coherent coupling with $\eta\gg 1$ between single neutral atoms and photons has been possible for over 20 years by using short cavities between one millimeter and a few tens of micrometers long \cite{Thompson1992,Colombe2007}, this regime has been elusive for single ions due to the poor compatibility with dielectrics. Nonetheless, a number of experiments with centimeter-sized cavities have pushed the single-ion cooperativity to a regime $\eta\sim 1$ \cite{Mundt2002,Keller2004,Barros2009,Stute2012a}, demonstrating single-photon sources \cite{Keller2004,Barros2009} and deterministic ion-photon entanglement \cite{Stute2012a}. One approach to get stronger coherent coupling is to reduce the cavity waist by going to sub-millimeter cavities and trying to mitigate the dielectric charging effects \cite{Steiner2013,Herskind2011}. Another approach is to couple the cavity collectively to a large number of ions ($N$), thus enhancing the cooperativity $\eta$ by the factor $N$. Strong collective coupling has been achieved this way with three-dimensional Coulomb crystals \cite{Herskind2009a,Albert2011}. In this subsection, we describe our system for collective ion-photon coupling with linear ion chains \cite{Cetina2013}, suitable for motional quantum gates with ions. We also present some other applications of cavities in ion traps in the intermediate-coupling regime, ranging from motional spectroscopy to spectral engineering.

\begin{figure}
\includegraphics[width=3.0in]{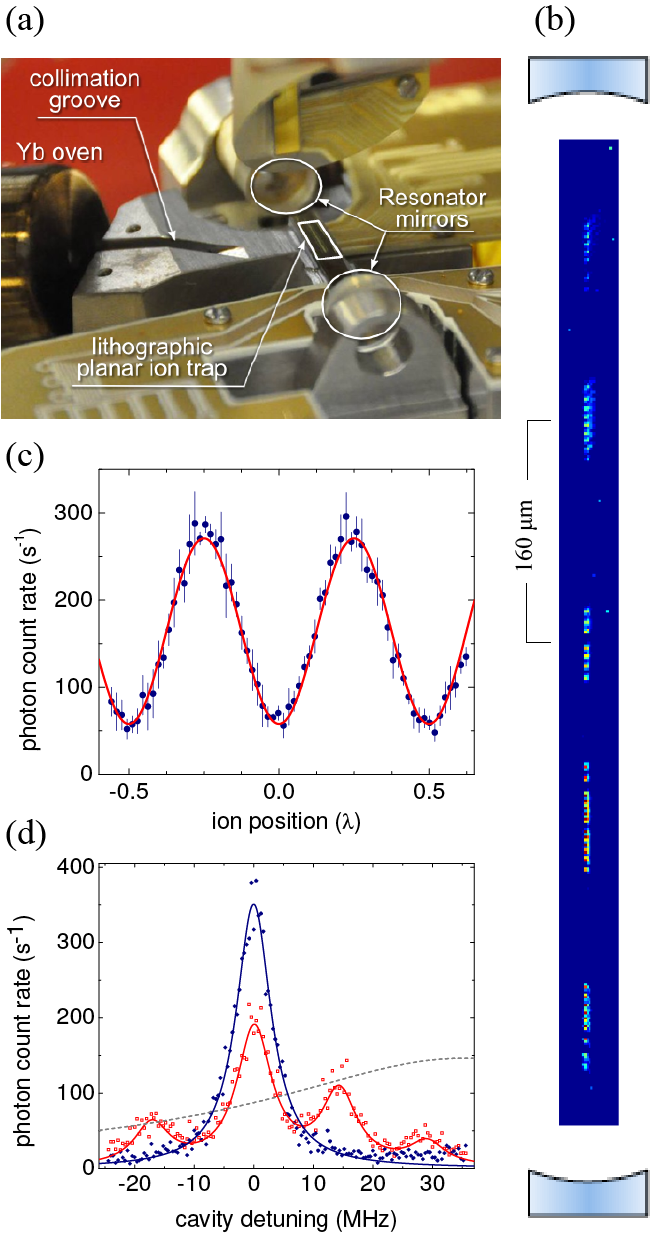}
\caption{A microfabricated planar trap array of ion chains coupled to an optical cavity forms a collective ion-photon interface. Reproduced from Ref.~\cite{Cetina2013}. (a) Photograph of the integrated microfabricated ion trap and optical cavity (length 22 mm) inside the vacuum chamber. (b) Photograph of 5 microtraps of the array loaded with ion chains. The cavity mirrors (not to scale) are shown to illustrate array alignment with the cavity; 50 microtraps overlap with the mode in reality. (c) The fluorescence from a single ion collected by the cavity is modulated by the ion's position with respect to the cavity with a period of half the wavelength. (d) Cavity as a spectrometer: micromotion sidebands are resolved (red) and can be compensated away (blue).}
\label{fig:IonArray}
\end{figure}

Cetina et al.~\cite{Cetina2013} created a versatile ion-cavity system, integrating a planar microfabricated ion trap with a medium-finesse optical cavity (see Fig.~\ref{fig:IonArray}a). The system is capable of strong coherent coupling between collective spin states of multiple chains of Yb\textsuperscript{+} ions and the cavity mode, while preserving the linear structure of the chains that lends itself to the trapped-ion quantum computing toolbox \cite{Home2009}. This is enabled by an innovative microfabricated trap architecture with an 8~mm-long trapping region, broken into a periodic array of 50 microtraps, each holding a short chain of ~10 ions (see Fig.~\ref{fig:IonArray}b). This trapping region at 138~$\mu$m from the microfabricated surface is carefully aligned to overlap with the mode of the near-confocal optical cavity with a waist of 38~$\mu$m and length of 2.2~cm, thus keeping the dielectric cavity mirrors sufficiently far from the ions to mitigate unwanted potentials, while maximizing the $N$-fold enhancement of the cooperativity, achieving $\eta \approx 0.2$.

Cetina et al.~\cite{Cetina2013} demonstrated coupling of the ions to the cavity via collection of fluorescence photons with the expected collection efficiency. Precise control of the coupling via nanopositioning of ions with respect to the longitudinal cavity mode was demonstrated as shown in Fig.~\ref{fig:IonArray}c. Micromotion fluorescence spectroscopy was performed on the ions by tuning the cavity resonance across the sideband spectrum (Fig.~\ref{fig:IonArray}d), demonstrating the cavity's functions beyond those of simple collection optics: as a spectrometer and for photon confinement. Loading of the microtraps with deterministic ion numbers, also important for controlled collective coupling, was demonstrated and was limited only by \textless 10\% residual isotopic impurities, which can be virtually eliminated by loading from a built-in magneto-optical trap (as demonstrated in \cite{Cetina2007} and \cite{Sage2012}, see Section \ref{sec:Hybrid Systems with Ions and Neutral Atoms}).

About 10 ions, easily confined as a small linear chain in each microtrap, are sufficient to reach strong collective coupling in this system, opening the possibility for entangling multiple separated ion chains via the common cavity mode. Coupling collective spin states to individual ions' spin states via shared vibrational modes (as proposed, for example, in \cite{Lamata2011a}), could then make it possible to scale up trapped-ion QIP from a few qubits, currently limited by the requirement to resolve the growing number of shared vibrational modes, to a few tens of qubits.

In addition to enhancing ion-photon coupling, a high finesse cavity can also significantly change the scattering spectrum of trapped ions. This can be a convenient tool for engineering the laser-atom interactions required for cooling and read-out in QIP. For instance, an optical cavity can be used to cool the motional state of an ion without affecting its internal (qubit) state, using a technique known as cavity cooling. Compared to ions trapped in free space, where the scattering spectrum is centered at the incident laser frequency, ions in a high finesse optical cavity can have a significant portion of their scattered photons emitted into a cavity mode. As a result, with the cavity resonance blue-detuned to the incident laser, cooling can be achieved as the ion preferentially emits higher energy photons into the cavity \cite{Horak1997,Vuletic2000}. Unlike conventional Doppler cooling techniques, involving spontaneous atomic dipole emissions \cite{Hansch1975}, cavity cooling can preserve the ion internal (qubit) state because the coupling mechanism allows for cooling lasers to be far-detuned from any atomic resonance. This is useful for trapped-ion QIP, where cooling the ions' motional degrees of freedom is required while preserving the coherence of internal states involved in information storage.

\begin{figure}
\includegraphics[width=2.6in]{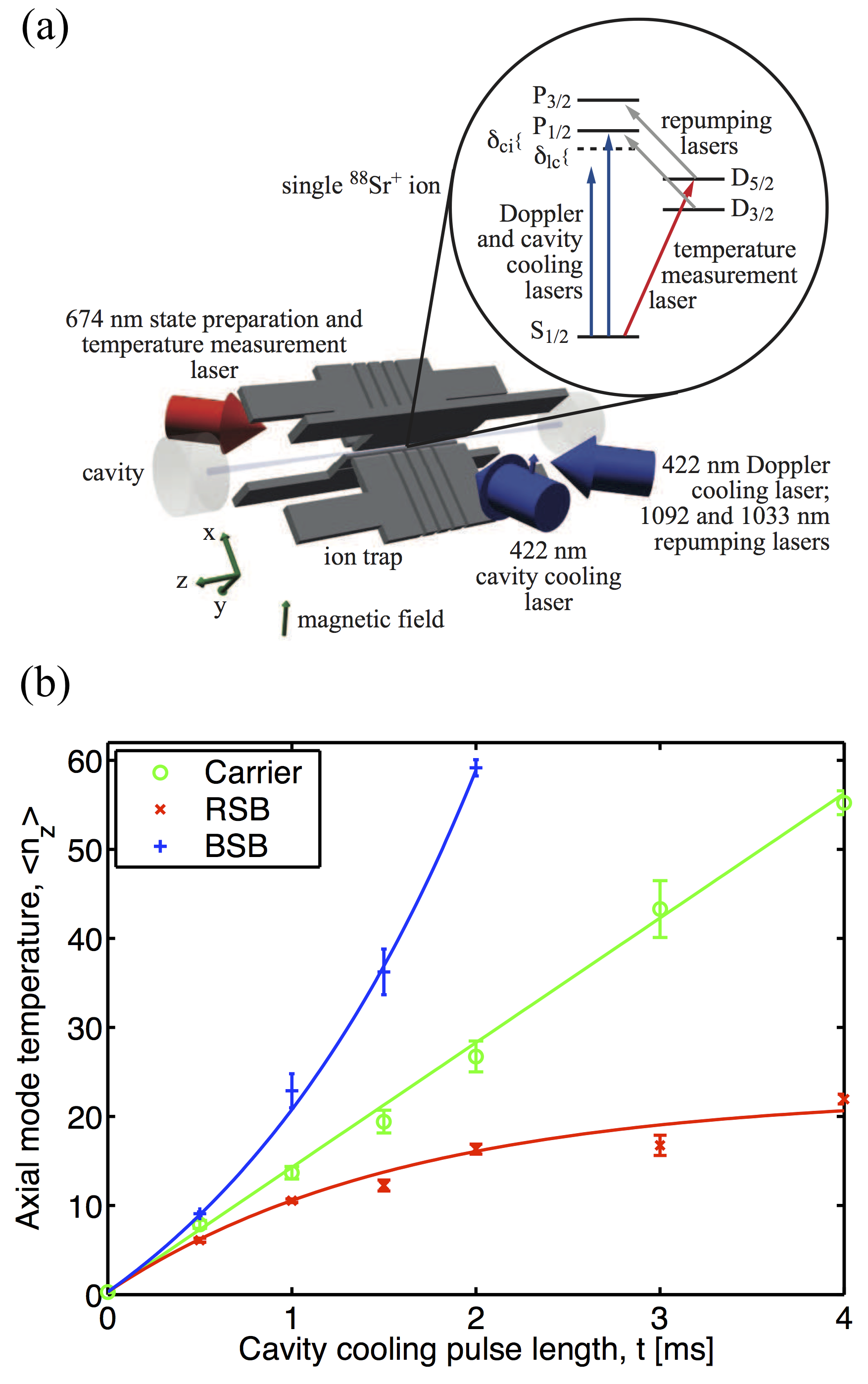}
\caption{Cavity cooling with a single ion in the resolved sideband regime has been demonstrated. Reproduced from Ref.~\cite{Leibrandt2009}. (a) Schematic of the experimental apparatus. A single $^{88}$Sr$^+$ is trapped (with secular frequency $\omega$) in the center of a linear RF Paul trap, and coupled to an optical resonator oriented along the trap RF-free ($z$) axis. The ion energy levels and cavity resonance are shown in the inset. (b) Plot showing the measured cavity cooling dynamics. After the ion is sideband cooled to the three-dimensional motional ground state, a cavity cooling laser is applied to the ion, with either frequency detuning $\delta_\mathrm{lc} = 0$ (carrier), $\delta_\mathrm{lc} = -\omega_z$ (red axial sideband), or $\delta_\mathrm{lc} = +\omega_z$ (blue axial sideband) to the cavity resonance. The mean number of motional quanta $\langle \mathrm{n}_z \rangle$ in trap $z$-axis is then measured. The relative motional heating rates observed for the three frequency detunings show a signature of cavity cooling.}
\label{fig:CavityCooling}
\end{figure}

Cavity cooling was proposed theoretically by Vuleti\'{c} et al.~\cite{Vuletic2000,Vuletic2001} and was first demonstrated with neutral atoms~\cite{Maunz2004,Nubmann2005,Fortier2007,Chan2003,Black2003}. Leibrandt et al.~\cite{Leibrandt2009} realized cavity cooling with a single ion in the resolved sideband regime, where the cavity linewidth $\kappa$ satisfies $\kappa\ll \omega$, the ion secular frequency. In this setting, the cavity resonance is blue-detuned by $\omega$ to the incident laser (Fig.~\ref{fig:CavityCooling}), and because of the sufficient frequency discrimination provided by the cavity, ground state cooling is possible. Fig.~\ref{fig:CavityCooling} shows how motional heating is altered by changing the laser detuning with respect to the cavity resonance. It shows that the heating rate is clearly suppressed in the presence of the cavity cooling (red detuned) laser, and is exacerbated in the presence of a blue detuned laser, matching the theory for cavity cooling.

\begin{figure}
\includegraphics[width=2.8in]{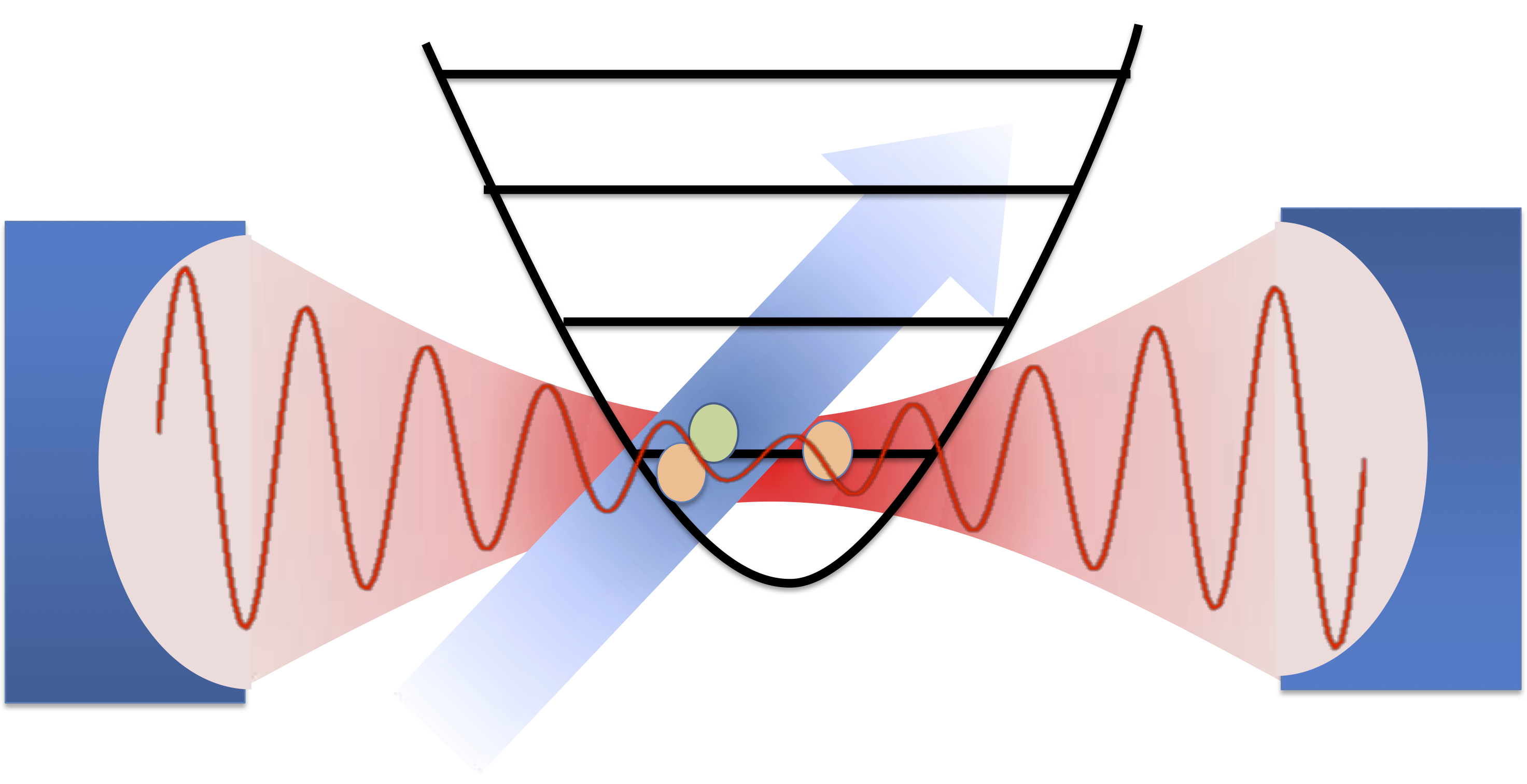}
\caption{A scheme to create a non-zero effective Lamb-Dicke parameter has been developed which enables direct microwave addressing of molecular rotational states. A cavity standing wave generates a large electric field gradient, and thus a large gradient in the differential AC Stark shift between the molecular rotational states. This in turn enhances the rotational state-to-motional state coupling of the molecule. By co-trapping an atomic ion with a molecule, quantum logic spectroscopy using microwave addressing becomes possible, and ground state cooling of the molecular rotational states can be implemented~\cite{Shi2013}.}
\label{fig:MQLSSchematics}
\end{figure}

Another application of cavity fields lies in engineering the coupling between motional and internal states of an ion, which is useful for many quantum gate operations~\cite{Cirac1995}. Shi et al.~\cite{Shi2013} proposed a technique to widely tune the Lamb Dicke parameter of trapped ions in an optical cavity. When an electric field causes a differential AC stark shift between the ion internal states, a spatial variance of its strength can significantly alter the states' energy separation depending on the ion position. This results in an effective ion motional-internal state coupling, as measured by an effective Lamb Dicke parameter $\eta_\mathrm{eff}$. In a cavity integrated with an ion trap, this large electric field gradient can be realized by positioning the ions at the node of the cavity standing field. Although the engineering of $\eta_\mathrm{eff}$ has been achieved using magnetic field gradients~\cite{Mintert2001,Ospelkaus2008,Ospelkaus2011,Timoney2011}, the advantage of using electric fields is that this approach can be extended to internal states with nearly negligible magnetic moments, including rotational states of typical heteronuclear diatomic molecules. This is of interest for QIP because such states have exceptionally long coherence times, and so have potential to serve as a quantum memory. In particular, Shi et al. showed theoretically that by co-trapping a molecule and an atomic ion (with well-characterized cooling and qubit transitions), ground state cooling and quantum logic spectroscopy of the molecular rotational states is possible (see Fig.~\ref{fig:MQLSSchematics}).

\FloatBarrier
\section{A versatile RF-free trap: ions in an optical lattice}
\label{sec:A versatile RF-free trap: ions in an optical lattice}

Efforts to scale-up quantum information processors with ions in Paul traps have been limited in part by micromotion-related heating \cite{Berkeland1998}, the difficulty of crafting arbitrary multi-dimensional RF potentials on a small scale \cite{Metodi2005}, and the anomalous heating associated with ions close to surfaces (see section 2). An alternative means to trapping atomic ions is to exploit their electric susceptibility and confine them in a trap formed by an optical potential, identically to techniques developed for trapping neutral atoms \cite{Grimm2000}. Although dipole forces generated by typically achievable optical traps are much smaller than electric forces in a Paul Trap, the advantages of a micromotion-free environment, combined with the versatility of optical potentials in, for example, generating arrays of trapping sites \cite{Cirac2000}, have motivated a set of experiments aimed at optically trapping ions \cite{Schneider2010,Linnet2012,Karpa2013}. Assuming ions can be optically trapped long enough for a computational cycle, hybrid trapping systems combining electrostatic and optical potentials could offer an alternative to Paul traps for large-scale ion processors. Though far from achieving this goal, the systems presented in this section are stepping-stone experiments towards evaluating its feasibility.

Much progress has been made in recent years, with proof-of-concept demonstrations of optically trapping a single ion in hybrid systems composed of a single beam dipole trap \cite{Schneider2010,Huber2014}, or a 1D optical lattice \cite{Linnet2012,Karpa2013} with transverse confinement from a Paul trap and longitudinal electrostatic fields superimposed. All-optical trapping in a 1D optical lattice was also achieved, though with trapping lifetimes limited to 100~$\mu$s\cite{Enderlein2012}. More recently, single-optical-site trapping lifetimes upwards of 10~ms, aided by lower temperatures, have been demonstrated in the hybrid 1D optical lattice -- RF Paul trap \cite{Karpa2013}, described here. Ideally, however, lifetimes of many minutes would be achievable for a single ion, giving ample time for deterministic computation on a large ensemble of ions.

Lifetimes in these traps have been limited so far by heating in several ways. The deep optical potentials required to confine a charged particle against stray fields impart significant recoil heating \cite{Schneider2010,Huber2014}, especially when a relatively small detuning to atomic resonance is required due to limited laser power. The large optical fields required also cause significant light shifts on the atomic transitions, limiting the cooling efficiency of typical techniques, such as Doppler cooling \cite{Karpa2013}, and to a lesser degree resolved-sideband cooling. These trapping fields can also be responsible for parametric heating from intensity fluctuations \cite{Savard1997}. Lastly, configurations where a 1D optical lattice has overlap with an RF Paul trap have suffered from off-resonant RF heating \cite{Enderlein2012}. This RF heating would not be present in hybrid systems combining only electrostatic and optical potentials.

\begin{figure}
\includegraphics[width=3.2in]{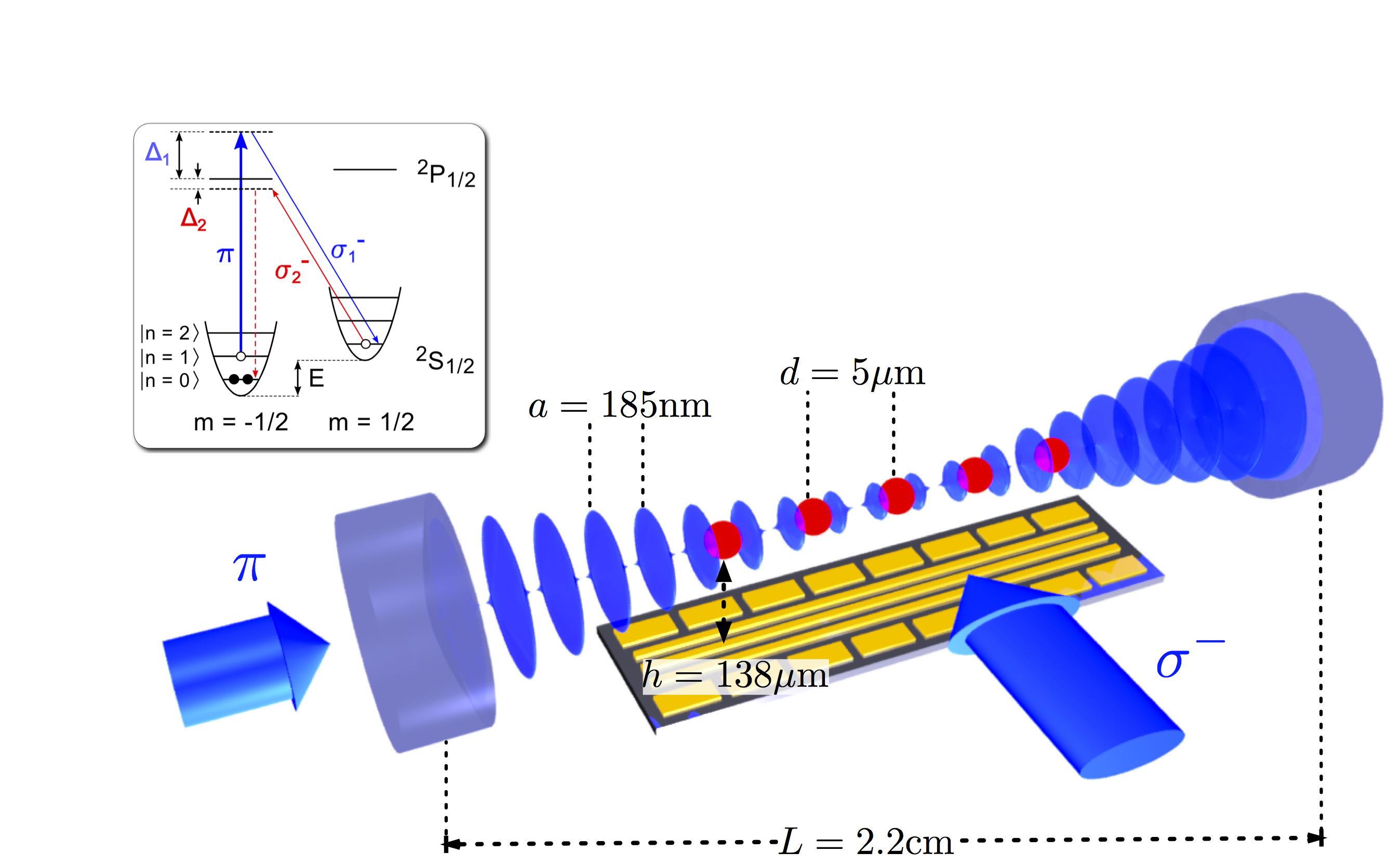}
\caption{A hybrid 1D optical lattice - RF Paul trap system (pictured here) has been developed \cite{Karpa2013,Cetina2013}. The dimensions shown are the optical cavity length L, the ion height relative to the trap h, the typical ion-ion distance d, and the lattice site spacing a. The optical lattice is $\Delta_{1} = 12$~GHz detuned to the blue of the atomic S-P transition in $^{174}$Yb$^+$, with a wavelength of 369.5~nm. Typical lattice depths are $h \times 45$~MHz, corresponding to secular frequencies in each optical lattice site of $\approx 2 \pi \times 1.2$~MHz. The axial electrostatic confinement can be tuned from $2 \pi \times 20$~kHz, to $2 \pi \times 1$~MHz. The $\pi$ polarized lattice light, coherent with a $\sigma^{-}$ polarized side-beam ($\sigma^{-}_{1}$), are used in a Raman Sideband cooling scheme. A near-detuned $\sigma^{-}$ beam ($\sigma^{-}_{2}$), collinear with $\sigma^{-}_{1}$, provides the optical pumping and dissipation channel.}
\label{fig:Setup_5ions_labels}
\end{figure}

In a hybrid system combining a Paul trap and a 1D optical lattice, RF heating can be reduced by aligning the RF-free axis of the trap with the optical lattice. Issues related to light shifts can be mitigated with a blue-detuned lattice, which traps at intensity minima, while trap intensity fluctuations can also be reduced by generating the lattice with a narrow optical cavity (as long as the trap laser's spectral width is narrower than the cavity's). The hybrid system developed by Karpa et al.~\cite{Karpa2013,Cetina2013}, combines these three advantages. It is composed of a planar Paul trap for radial RF confinement in 2 dimensions and for weak electrostatic confinement in the third dimension, and of a blue-detuned optical lattice formed by an optical cavity aligned with the RF-free axis (see Fig.~\ref{fig:Setup_5ions_labels}). The electrostatic trap provides the overall depth needed to maintain the ion chain in the Paul trap, while the low-depth optical lattice confines each ion tightly in separate lattice sites.

In addition to providing a periodic trapping potential, the optical lattice is used in a Raman sideband cooling scheme that couples the two magnetic sub-levels of $^{174}$Yb$^+$'s electronic ground state \cite{Karpa2013}, as in Fig.~\ref{fig:Setup_5ions_labels}. A circularly polarized beam near-detuned to the red is addressed at the ions perpendicular to the lattice, and provides the optical pumping and tunable dissipation mechanism needed to complete the cooling cycle (at a scattering rate $\Gamma_{sc} \approx 2 \times 10^5$ s$^{-1}$). By virtue of its spatially-dependent coupling by the optical lattice, this cooling scheme also gives rise to a spatially-dependent fluorescence, used in measuring an ion's temperature and average position relative to the optical potential \cite{Bylinskii2015}. In this system, temperatures as low as $\langle n \rangle = 0.1$ have been realized with a single ion (for an optical lattice vibrational frequency of $\omega/2\pi = 1.2$~MHz and depth $U/h = 45$~MHz). From a simple Arrhenius rate calculation where the lifetime $\tau$ is approximately $1/\Gamma_{sc} \exp(U/\langle n \rangle \hbar \omega)$, this temperature corresponds to lifetimes under continuous cooling of many minutes in a single lattice site at the deepest point of the potential. Ion crystals of up to 3 ions, arranged in such a way that each ion's average position corresponds to a lattice minimum, have been cooled to similar temperatures.

\begin{figure}
\includegraphics[width=3.2in]{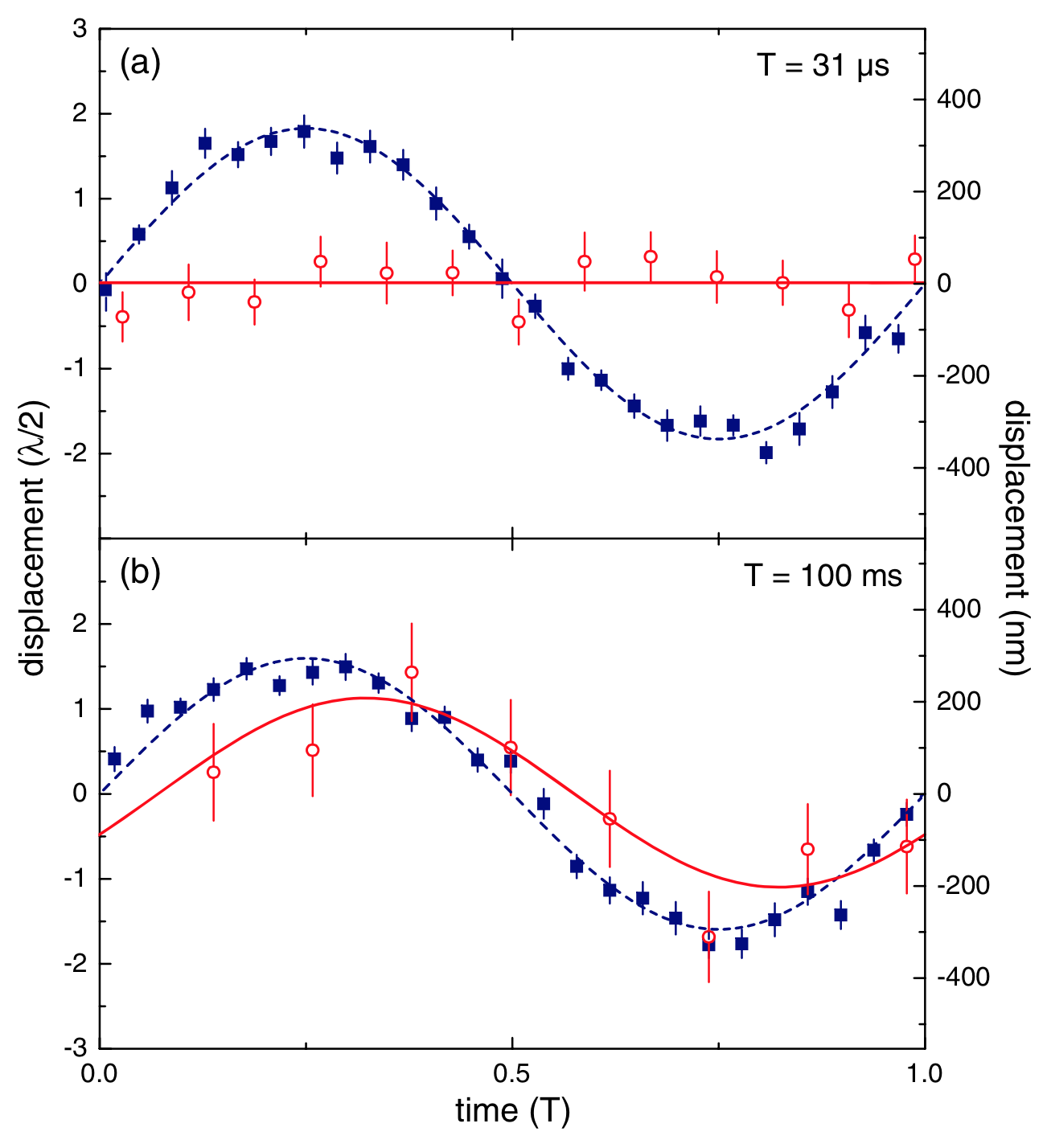}
\caption{The lifetime of a single ion in an optical lattice site of the hybrid 1D optical lattice - RF Paul trap system has been measured. Reproduced from Ref.~\cite{Karpa2013}. A sinusoidal electrical drive is applied to the ion to move it back and forth along the lattice direction. The time axis is in units of the drive period. In blue, the position versus time in the absence of the lattice, while in red, the position versus time in the presence of the lattice. (a) for a drive period of 31~$\mu$s, the motion is completely suppressed by the lattice, showing that the ion remains in a single site on that time scale. (b) for a drive period of 100~ms, the motion is partially suppressed by the lattice; for a fraction of the trials, the ion has time to thermally hop to adjacent sites (with a direction biased by the drive).}
\label{fig:Lattice-data}
\end{figure}

Experimentally, Karpa et al.~\cite{Karpa2013} determined the ion lifetime in a single site by measuring its position on a quadrant-type detector, and inferring the hopping probability on a range of time scales. By electrically driving the ion back and forth and measuring its position, the confinement by the optical lattice could be readily seen. If the drive's oscillation period was low enough, the ion would complete many driven, small amplitude oscillations in the site before thermally hopping to an adjacent site; in this case the motion averaged to nearly zero over the drive period (see Fig.~\ref{fig:Lattice-data}a). As the drive period increased, the ion had progressively more time to thermally leave its initial site, thus giving rise to a large driven oscillation in position over the drive period (see Fig.~\ref{fig:Lattice-data}b). From this perturbation-based experiment, Karpa et al.~\cite{Karpa2013} measured lifetimes in a single lattice site upwards of 10~ms, setting a very conservative lower bound on the lifetime in the absence of perturbation.

This experiment was conducted under continuous cooling, however, and as such some challenges remain before even simple quantum operations can be performed in such an optical trap with ions. Indeed, anomalous heating rates of one to a few vibrational quanta per ms have been measured in this system. While typical for a room-temperature surface trap of these dimensions, they set a prohibitive time scale for coherent manipulation of the ion's motional degrees of freedom. As a future direction for this type of hybrid system, lower heating rates can be achieved by placing the ion trapping region farther away from electrodes, as is typically the case in a conventional four-blade Paul trap (see section 2).

While optical traps for ions introduce their own sets of challenges in comparison to conventional Paul traps, such as new heating mechanisms and limited trap depths, they offer a radical technological alternative for ion-based quantum computation well worth further investigation. In particular, promising theoretical estimates of second-scale single-site lifetimes (based on measured temperatures), combined with the potential ability to replace the RF Paul trap with a purely electrostatic confining quadrupole configuration in the system described above, open the possibility of a versatile multi-site, micromotion-free ion trap, where further development on tackling heating rates could yield a new viable path for an ion-based quantum computer.

\section{Hybrid Systems with Ions and Neutral Atoms}
\label{sec:Hybrid Systems with Ions and Neutral Atoms}

Large scale quantum computation with ions will require large numbers of ions trapped simultaneously \cite{Cirac2000}. This will presumably entail complicated surface trap structures, where arrayed trapping sites are used as registers, demanding, for fast high-fidelity operation, close neighboring sites and tight traps. Loading such traps with standard thermal beam photoionization presents a number of problems \cite{Sage2012}. To enable trap miniaturization, trap depths will be limited to levels well below the capture range for a thermal ion distribution, thus vastly limiting the loading rate. The large fraction of hot metallic atoms unaffected by the ionization process have a significant chance of coating electrodes and dielectric forming the trap, thus causing shorting and additional noise, particularly in architectures involving close integration. In addition, atoms photoionized from a hot vapor, which have limited isotope-selectivity, will populate the trap with unacceptable levels of defect ions. While methods such as backside loading \cite{Britton2009,TrueMerrill2011} circumvent the problems associated with spurrious coating, it is unclear whether even the use of highly pure single-isotope materials as a source can reach the rates of isotopically pure loading required to populate and operate large arrays of ions. One way to address all three problems is to use a cold, isotope-selective \cite{Chen1999}, stationary source of atoms: a Magneto-Optical Trap (MOT). Though adding some engineering overhead, the MOT is a mature atom-trapping platform \cite{Rushton2014} that has been integrated with ion-traps without excessive technological difficulties \cite{Cetina2007,Cetina2012,Sage2012}.

\begin{figure}
\includegraphics[width=3.2in]{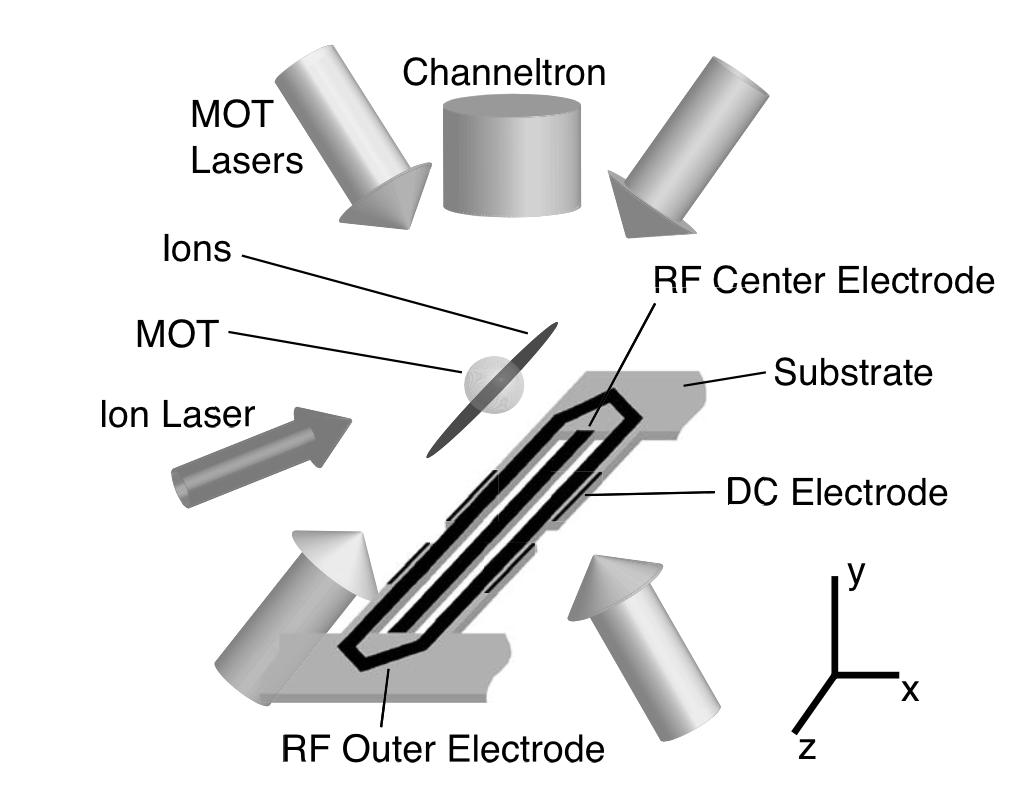}
\caption{An apparatus for loading a surface-electrode ion trap by in-trap photoionization of laser-cooled atoms has been developed. The setup is outlined here. Reproduced from Ref.~\cite{Cetina2007}. A Yb MOT is formed 4~mm above the trap surface. Cold Yb ions are produced inside the Paul trap by single-photon excitation from the excited $^1$P$_1$ atomic state.}
\label{fig:MOTIons}
\end{figure}

The experiments of Cetina et al.~\cite{Cetina2007} and Sage et al.~\cite{Sage2012} have both demonstrated loading of a surface trap from a MOT. Cetina et al.~\cite{Cetina2007} overlapped cold Yb atoms in a MOT with the trapping region of an ion trap and photoionized (see Fig.~\ref{fig:MOTIons}), resulting in loading rates of up to $4 \times10^5$~ions/second in a 0.4~eV trap, with direct loading persisting down to a trap depth of 0.13~eV. Sage et al.~\cite{Sage2012} used a resonant laser beam to push cold Sr atoms in a MOT (loaded from a remote oven with no line-of-sight to the ion trap) into the ion-trapping region of a cryogenic surface trap, where they are photoionized. There, loading rates of up to 125~ions/s have been reached for a trap depth as low as 0.015~eV. 

Beyond the important question of efficient ion-trap loading, an intriguing application of a hybrid system containing trapped atoms overlapped with trapped ions could be to exploit the ultra-low temperatures achievable by superfluid atoms in a Bose-Einstein condensate for the continuous sympathetic cooling of an ion-based quantum computer \cite{Daley2004,Zipkes2010b}. Another, less developed, potential application is the use of atom-ion collisions as a means to perform quantum gates \cite{Idziaszek2007a}. Neutral atoms are readily loaded with high packing factors into the wavelength scale traps of optical lattices, where they are long-lived and mostly free of environmental noise \cite{Jaksch2005}, and where single-site resolution is possible \cite{Bakr2009}. Exploiting this atom-ion interaction could lead to architectures where one or more ions are used as transportable quantum read/write heads to mediate interactions between neutral atoms forming a large quantum processor \cite{Doerk2010a}.

Towards this goal, an experimental observation of cold collisions between trapped ions and atoms in a MOT yielded a confirmation of a semi-classical prediction for the interaction rates \cite{Grier2009}. Subsequent theoretical work, however, put stringent bounds on the possibility of controlled quantum collisions between atoms and RF-trapped ions \cite{Cetina2012}. Indeed, the induced dipole interaction of the atom with the ion displaces the ion from the RF null, causing it to undergo micromotion; the subsequent collision with the atom causes an interruption of this micromotion during which energy from the RF drive can be converted to incoherent kinetic energy, and, on average, leads to heating rates generally prohibitive to reaching the quantum regime of atom-ion interactions. Cetina et al. \cite{Cetina2012} showed that the regime of single-partial-wave atom-ion interaction could only be reached for large ion-atom mass ratio (such as Yb$^+$ and Li), weak RF traps, and state-of-the-art control over DC electric fields on the order of 10~mV/m (as recently achieved in \cite{Huber2014}).

Despite these limitations, the technological developments presented here, combining atoms and ions, highlight important progress and new directions towards a scalable architecture accomplished by hybridizing ion traps.

\section{A quantum system built with classical computer parts}
\label{sec:A quantum system built with classical computer parts}

In the systems described thus far, traps were constructed using specialized, non-standard fabrication processes. The benefit of specialized fabrication processes---which are the norm in the ion-trapping community---is that they allow each trap to be tailored to the materials or devices to be integrated. The downside of using custom processes is that repeatability at different facilities is very difficult, and features like the lithographic resolution are typically limited. All of this is in contrast to how classical processors are manufactured, using standardized CMOS (complementary metal-oxide-semiconductor) technology, which allows for devices with critical dimension of $\approx10$~nm, in quantities of about $10^{11}$, to be fabricated reproducibly on a common silicon substrate. A typical ion trap fabrication process involves patterning a single metal layer atop a quartz or sapphire substrate.

By contrast, a standard, high-resolution CMOS fabrication process typically involves patterning of 8 or more copper interconnect layers (separated by oxide) plus an aluminum pad layer, on a doped polysilicon substrate; the resulting active and passive layers can be used to form dense, high-performance digital circuits. Silicon substrates have been used in trap construction before, but none of these traps have had doped, active device fabrication available, and only a few metal layers (four maximum) have been implemented to date \cite{Leibrandt2009b,Stick2010,Britton2008,Allcock2011b,Wilpers2012,Wright2013,Sterling2014,Niedermayr2014}.

To probe whether the performance and scale of the CMOS manufacturing platform could be leveraged for trapped-ion systems, Mehta et al.~\cite{Mehta2014} designed a surface-electrode ion trap to be fabricated in a commercial CMOS foundry process, and demonstrated trapping.

Some of the challenges anticipated in creating a CMOS trap were: dielectric breakdown due to the large voltages needed for trapping, laser-induced charging of exposed silicon oxide dielectrics, and RF loss in the p-type doped silicon substrate. Fortunately, Mehta et al.~\cite{Mehta2014} found the trap could sustain voltages significantly higher than those used in CMOS electronics (the leakage current remained below 10~pA for up to 200~V static bias applied). But to address the other challenges, it was essential to include a ground plane between the trap electrode layer and the polysilicon substrate.

\begin{figure}
\includegraphics[width=3.3in]{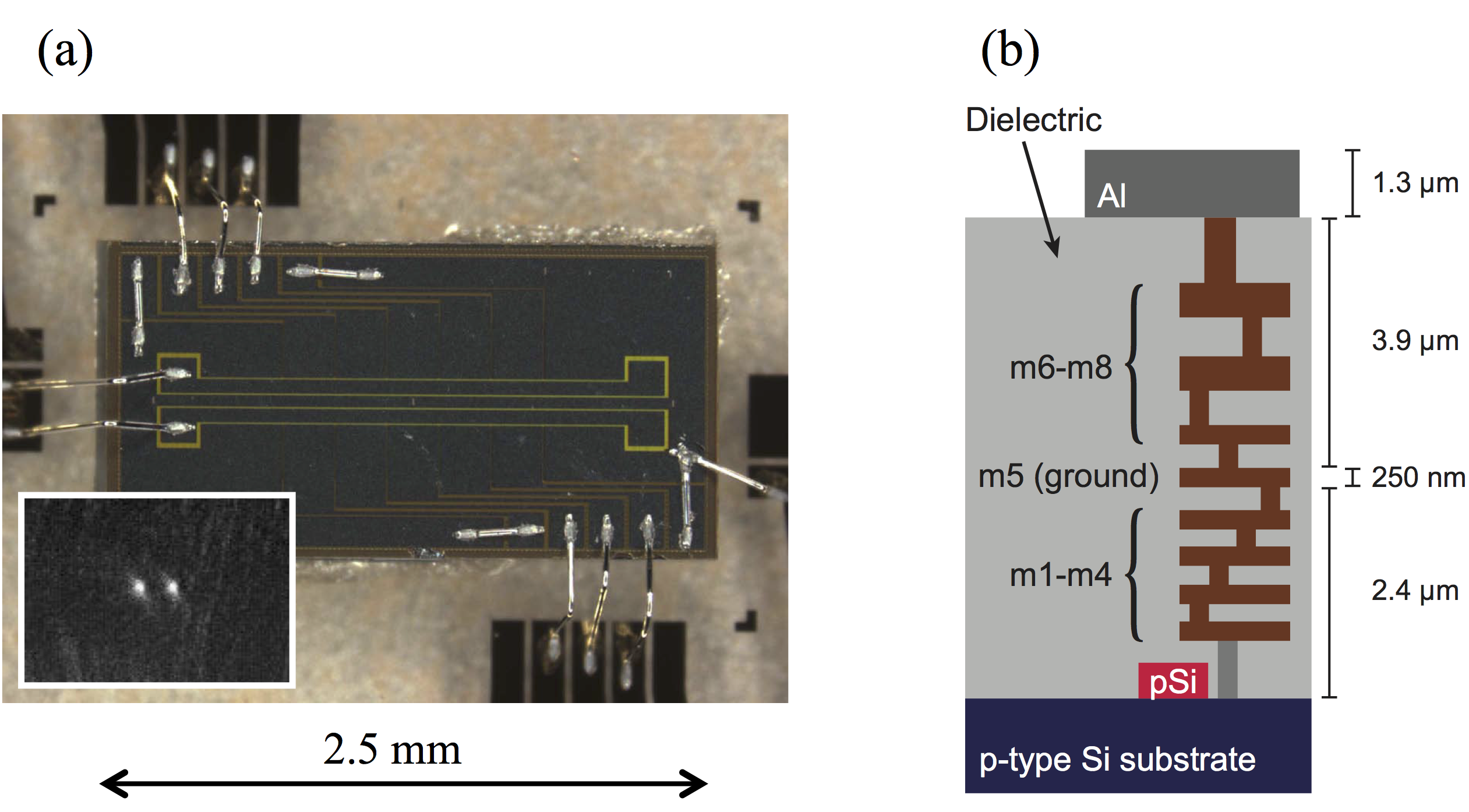}
\caption{A surface-electrode ion trap was fabricated entirely within a commercial CMOS foundry process. Adapted from Ref.~\cite{Mehta2014}. (a) Micrograph of trap chip diced from die and mounted on the sapphire interposer of a cryogenic vacuum system. Aluminum wirebonds are used to make contact from the aluminum trap electrodes to the gold interposer leads. The chip is $2.5$~mm long and $1.2$~mm wide. The inset shows two ions trapped 50~$\mu$m from the surface of the trap chip. The ions are approximately 5~$\mu$m apart. (b) A diagram of the chip cross section, with approximate relevant dimensions labeled (``pSi'' is polysilicon and metal interconnect layers are labelled m1 through m8). Vias shown between metal layers are only representative. The trap electrodes were formed in the top aluminum pad layer, and the ground plane was formed in metal layer m5.}
\label{fig:CMOS-figure}
\end{figure}

\begin{figure}
\includegraphics[width=2.8in]{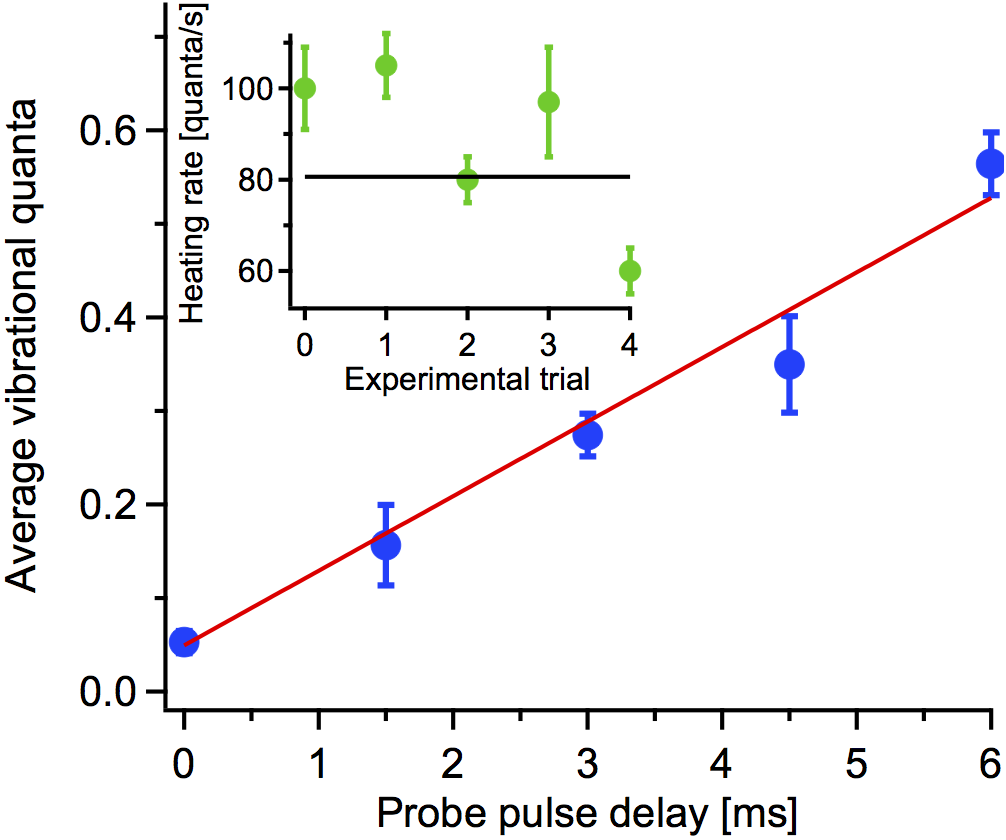}
\caption{The heating rate in a CMOS-foundry-fabricated ion trap has been measured. Plot shows average occupation of the axial mode of vibration in the linear trap as a function of delay time after preparation in the ground state at a trap temperature of 8.4~K. A linear fit (line shown) gives a heating rate for these data of 80(5)~quanta/s where the uncertainty is due to statistical errors propagated through the fit. An average of five such measurements gives a heating rate of 81(9)~quanta/s where the uncertainty is due to run-to-run variability. The inset shows all five measurements with a line indicating the weighted average value. The average motional heating rate is comparable to those seen in surface-electrode traps of similar size. Reproduced from Ref.~\cite{Mehta2014}.}
\label{fig:CMOS-heating}
\end{figure}

Devices were fabricated on a 200-mm wafer (shared with other projects) produced in a 90-nm CMOS process operated by IBM (9LP process designation) \cite{Mehta2014}. The $1.3$~$\mu$m thick aluminum pad layer (on top) was patterned to form the trap electrodes (see Fig.~\ref{fig:CMOS-figure}). Without the ground plane, the lasers used for Doppler cooling and photoionization induced photo-effects in the polysilicon substrate which interfered with the RF trapping potential. With a ground plane formed in a copper layer about 4~$\mu$m below the electrodes (and 2~$\mu$m above the substrate), RF leakage into the polysilicon substrate was reduced enough that ion trapping was stable and the motional heating rate measured was comparable to those seen in surface-electrode traps of similar size (see Fig.~\ref{fig:CMOS-heating}). Some technical issues remained, including slightly higher RF power dissipation in the trap, and increased light scattering from the electrodes (likely due to higher as-deposited roughness of the aluminum layer, relative to that of a standard, single-metal-layer trap).

This demonstration of trap hardware utilizing a fabrication process, without modification, that has enabled scaling to billions of transistors, opens the door to integration of CMOS-based electronics and photonics \cite{Field2010,Orcutt2012} for ion control and readout.  A large-scale, integrated optics architecture might have single-mode waveguides distributing light to various locations in a dielectric layer in the same chip as the trap electrodes, and focusing grating couplers directing the light, through gaps in the trap electrodes to $\mu$m-scale focuses at the ion locations. This approach could allow practical integration of many couplers, avalanche photodiodes for readout, electro-optic modulators, and control electronics into the same chip as the trap electrodes. It could potentially offer superior individual-ion addressing, efficient fluorescence read-out, and improved phase and pointing stability of control beams. Standardization of the foundry process also makes it possible for any group to repeatably produce identical devices.

\section{Conclusion}
\label{sec:Conclusion}

In summary, we have reviewed recent work at MIT developing and evaluating candidate technologies for trapped-ion QIP. Our work has focused on enhancing qubit coherence, and on integrating ions with naturally scalable systems. Towards improved qubit coherence, new ion trap electrode materials, and new coherent cooling and optical trapping technologies have been investigated. Towards a scalable quantum network implementation, hybrid ion trap systems incorporating optical elements, CMOS fabrication technologies, and trapped neutral atom clouds have been constructed.
 
Although at this stage, the search for new trap fabrication materials, such as graphene and superconductors, has not yet provided solutions to reduce ion motional decoherence, these investigations have shown how motional decoherence can be affected by surface contamination of ion traps. We believe that this approach of experimenting with material properties of ion traps holds promise for scalable ways to reduce motional decoherence. One idea would be to passivate the trap surface with a thin layer of a lipophobic material that prevents hydrocarbon molecules from sticking to the trap during the vacuum bake. Another approach would be to clean or chemically break down (crack) the contaminating hydrocarbon in-situ.

A spinoff of our superconducting ion trap work might be to couple trapped-ion qubits with superconducting qubits, thus combining the advantages of long internal coherence times of ions with fast gate speeds offered by microwave superconducting circuits. Similarly, there are several proposals to integrate polar molecules with superconducting devices, particularly for cavity QED experiments \cite{Andre2006,Schuster2011}. More generally, because of the possibility for resistanceless current flow, superconducting ion traps would likely prove useful for any trap integrating microwave waveguides, such as those employed for microwave near-field quantum control of the ion spin and motional degrees of freedom \cite{Mintert2001,Ospelkaus2011,Brown2011,Warring2013,Allcock2013}. Ground-state cooling using microwave radiation was recently demonstrated experimentally, which is advantageous because it removes the requirement for highly narrow laser sources or multiple laser beams in a Raman configuration \cite{Weidt2015}.

The proof-of-principle investigations into cavity cooling and optical lattice trapping open a door for alternative ways to improve motional coherence of ions: coherent laser cooling and micromotion-free (optical) micro-traps far from any surfaces. Coherent laser cooling could potentially be used to re-cool the motion of ions between ion transport and quantum gates without involving internal states or ions of different species for sympathetic cooling \cite{Home2009}. Another advantage of coherent laser cooling is that it does not rely on a particular internal level structure and is therefore portable to many different species of ions, atoms and even molecules, the latter potentially providing another promising physical qubit. Meanwhile, ion chains trapped in 1D and 2D optical lattices could provide a new type of scalable architecture for trapped-ion quantum computing, as well as enable quantum simulation schemes based on weakly-coupled oscillator networks \cite{Cirac2000,Schmied2008,Bermudez2011}, or simulations of crystal lattice mechanics \cite{Bylinskii2015}.

The demonstrated integration of optical elements with ion trap systems could pave the way for both probabilistic and deterministic implementations of ion-photon quantum interconnects that can link together a large quantum network. The nodes of such a network could be single ions or ion chains coherently coupled to separate optical cavities \cite{Ritter2012}. The cavity mirrors could be miniaturized by fabricating them into the trap, as we have shown, or on fiber tips \cite{Steiner2013}, providing the compact implementations needed for scalable quantum networks. In addition, arrays of fibers addressing different ions could be used to effectively deliver control and cooling beams, and ion fluorescence could be collected by the same fibers, or detected by detector arrays placed just below ion chains behind transparent trap electrodes.

The experimental demonstration of an ion trap fabricated using standard CMOS processes opens up a tantalizing opportunity to make use of the wide ranging capabilities of modern CMOS systems to push trapped-ion quantum information systems to next levels of scalability and compactness. One could also imagine making use of integrated silicon nanophotonic waveguides and grating couplers to produce quantum networks with ions and photonic interconnects on a chip \cite{Field2010,Orcutt2011}.

Although major progress has been made in evaluating a broad range of candidate technologies for scaling-up ion-based QIP, significant effort is still required to refine and combine these advancements, to make the systems compact and to push gate fidelities past error-correction thresholds. Having said that, we note that the versatility of planar ion trap systems has made it possible to incorporate advanced technologies from a wide range of fields, indicating great promise for further development. We hope that the work presented in this review article will encourage continued explorations into hybrid ion traps and developments of scalable quantum computing technologies.

\FloatBarrier
\begin{acknowledgements}
We gratefully acknowledge support from the MQCO Program with funding from IARPA, the Quest program with funding from DARPA, the Air Force Office of Scientific Research MURI on Ultracold Molecules, and the NSF Center for Ultracold Atoms. AME, DG, and AB also gratefully acknowledge the support of the National Science and Engineering Research Council of Canada's Postgraduate Scholarship program.
\end{acknowledgements}

\bibliographystyle{spphys}       
\bibliography{review}   

\end{document}